\documentclass[12pt]{article}
\usepackage{graphicx}

\begin{document}

\title{Quasiperiodicity in time evolution of the Bloch vector under the thermal Jaynes-Cummings model}

\author{Hiroo Azuma${}^{1,}$\thanks{On leave from
Information and Mathematical Science Laboratory Inc.,
Meikei Building, 1-5-21 Ohtsuka, Bunkyo-ku, Tokyo 112-0012, Japan.
Email: hiroo.azuma@m3.dion.ne.jp}
\ \ 
and
\ \ 
Masashi Ban${}^{2,}$\thanks{Email: m.ban@phys.ocha.ac.jp}
\\
\\
{\small ${}^{1}$Advanced Algorithm \& Systems Co., Ltd.,}\\
{\small 7F Ebisu-IS Building, 1-13-6 Ebisu, Shibuya-ku, Tokyo 150-0013, Japan}\\
{\small ${}^{2}$Graduate School of Humanities and Sciences, Ochanomizu University,}\\
{\small 2-1-1 Ohtsuka, Bunkyo-ku, Tokyo 112-8610, Japan}
}

\date{\today}

\maketitle

\begin{abstract}
We study a quasiperiodic structure in the time evolution of the Bloch vector,
whose dynamics is governed by the thermal Jaynes-Cummings model (JCM).
Putting the two-level atom into a certain pure state
and the cavity field into a mixed state in thermal equilibrium at initial time,
we let the whole system evolve according to the JCM Hamiltonian.
During this time evolution, motion of the Bloch vector
seems to be in disorder.
Because of the thermal photon distribution,
both a norm and a direction of the Bloch vector change hard at random.
In this paper, taking a different viewpoint compared with ones that we have been used to,
we investigate quasiperiodicity of the Bloch vector's trajectories.
Introducing the concept of the quasiperiodic motion,
we can explain the confused behaviour of the system as an intermediate state
between periodic and chaotic motions.
More specifically, we discuss the following two facts:
(1)
If we adjust the time interval $\Delta t$ properly,
figures consisting of plotted dots at the constant time interval acquire scale invariance
under replacement of
$\Delta t$ by $s\Delta t$,
where $s(>1)$ is an arbitrary real but not transcendental number.
(2)
We can compute values of the time variable $t$,
which let $|S_{z}(t)|$ (the absolute value of the $z$-component of the Bloch vector) be very small,
with the Diophantine approximation (a rational approximation of an irrational number).
\end{abstract}

\section{\label{section-introduction}
Introduction}
The Jaynes-Cummings model (JCM) describes a two-level atom interacting
with a single-mode radiation field.
Because it is a soluble fully quantum mechanical model,
the JCM has been attracting many researchers' attention
since it was proposed in 1963
\cite{Jaynes1963,Shore1993,Louisell1973,Barnett1997,Schleich2001}.
Considering the Hamiltonian for a magnetic dipole in a magnetic field,
assuming near resonance and applying the rotating-wave approximation to it,
we obtain the Hamiltonian of the JCM.
Thus, the JCM is considered to be suitable for describing interaction of the radiation with the matter
in the semiclassical approximation at first,
for example, the spontaneous emission
given rise to by a single two-level atom in the cavity field.

If we put the single-mode field into the state with sharply defined photon number at initial time,
the JCM causes the Rabi oscillations in the atomic population inversion.
Moreover, if we put the field into the coherent state initially,
the JCM shows the collapse and the revival of the Rabi oscillations.
Because the classical theory cannot explain the revival of the Rabi oscillations,
we can regard it as an evidence of the quantum nature of the radiation,
that is, the discreteness of photons
\cite{Cummings1965,Eberly1980,Rempe1987}.

However, in the case where we prepare the single-mode field in a thermal state initially,
we hardly predict how the system evolves.
Because of the thermal fluctuation of the field,
the Bloch vector
develops in time in a confusing manner.
Both its norm and direction change hard at random,
so that it seems to be in disorder.
A number of researchers analyse this problem in details
and try to find distinct properties, which characterize its unpredictable behaviour.

In Ref.~\cite{vonFoerster1975},
von Foerster computes time evolution of the probability for finding the atom
in the ground state with preparing the radiation field into various states at initial time.
Knight and Radmore investigate both the thermal JCM
and its corresponding semiclassical model
\cite{Knight1982}.
They compare time evolution of the atomic population inversion of the quantum model
with that of the semiclassical model by plotting their graphs numerically.
In Ref.~\cite{Knight1986},
Knight analyses the time dependence of the atomic inversion of the JCM
with an initial field state in thermal equilibrium.
In the work, he states the following observation:
The thermal fluctuation causes the collapse of the Rabi oscillations
and their collapse time is very short.
In addition, the revival overlaps it,
so that interference occurs and it generates a very irregular time evolution.
Liu and Tombesi discuss thermodynamics of the JCM
\cite{Liu1992}.
In their analysis, the grand partition function of both the atom and the boson field is considered.

When we discuss the thermal JCM,
we have to handle an intractable infinite series.
If the cavity field is resonant with the atom,
the $n$th term of this intractable infinite series is given by a trigonometric function of $\sqrt{n}t$
for $n=0, 1, 2, ...$,
where $t$ represents the variable of the time.
Buck and Sukumar emphasize this fact in Ref.~\cite{Buck1981}.
Because the infinite series is a superposition of the trigonometric function of $\sqrt{n}t$
for $n=0, 1, 2, ...$ and it cannot be a Fourier series,
the value of the sum of the series varies in an unpredictable manner
as the time $t$ progresses.
Arroyo-Correa and Sanchez-Mondragon try to discuss the thermal JCM
and evaluate the atomic population inversion,
which is described by this intractable infinite series,
using a technique of the complex analysis
\cite{Arroyo-Correa1990}.
Chumakov {\it et al}. examine a new analytical approach to obtain an approximate sum of this series,
which is reliable for a small initial mean photon number
\cite{Chumakov1993}.
In Ref.~\cite{Klimov1999},
Klimov and Chumakov rewrite this intractable infinite series as a sum of two integrals
by using the Abel-Plana formula.

Not only the thermal fluctuation but also a continuous measurement lets the JCM exhibit
chaotic behaviour.
Fukuo {\it et al}. study time evolution of non rotating-wave approximation JCM
under a continuous quantum-nondemolition (QND) measurement
\cite{Fukuo1998}.
Counting the number of photons inside the cavity with the QND photodetector,
the system evolves in an irreversible way and yields chaos.

In Ref.~\cite{Hioe1983},
Hioe {\it et al}. investigate the long-time behaviour of the JCM from a statistical point of view.
They find that  there exists a mean angular frequency
characterizing the oscillation of the atomic variables
in both cases when the initial photon distributions are given
by a coherent state and a thermal equilibrium state.
Utilizing the Kronecker-Weyl theorem to replace the time average with the phase average,
Hioe {\it et al}. evaluate the atomic inversion and derive its mean angular frequency.

In Ref.~\cite{Yoo1981},
Yoo {\it et al}. investigate long-time behaviour of collapse and revivals of the Rabi oscillations
in the JCM.
The revivals become less complete and begin to broaden,
eventually overlapping each other at still longer times.
The saddle point approach is performed to the atomic inversion
and interference between revival signals is examined.
This systematic analysis is related to the notion of fractional revivals,
which is pointed out in Ref.~\cite{Averbukh1989}.

In Ref.~\cite{Averbukh1989},
Averbukh and Perelman study the long-term evolution of quantum wave packets
and establish the concept of the fractional revivals.
They indicate that the fractional revivals can be found
in a detailed numerical study of the Rydberg wave packets \cite{Parker1986}.
The fractional revivals are also observed in the time evolution of the atomic inversion
for the JCM with initial coherent state of the cavity field.
As time passes,
the revivals of the atomic inversion become broader and their amplitude decreases.
Eventually the revivals overlap and new structures develop.
These new structures are regarded as the fractional revivals \cite{Schleich2001}.

The results obtained in Refs.~\cite{Hioe1983,Yoo1981}
suggest the quasiperiodicity of the JCM.

In this paper, we study the time evolution of the Bloch vector governed by the thermal JCM.
Preparing the atom and the single-mode field initially in a certain pure state and
a mixed state with a Bose-Einstein photon number distribution respectively,
we let the whole system develop according to the JCM Hamiltonian.
Thus, the system never suffers from dissipation.
As mentioned before, under these circumstances,
the behaviour of the Bloch vector is thrown into disorder.
In this paper, taking a different viewpoint compared with ones that we have been used to,
we investigate quasiperiodicity of the Bloch vector's trajectories.
Using the concept of the quasiperiodicity,
we can explain the confused behaviour of the system as an intermediate state
between periodic and chaotic motions.
The revealed quasiperiodic structures hidden in the thermal fluctuation of the Bloch vector have
close relations with topics of the number theory,
which is one of the oldest branches of the pure mathematics.
This appearance of unexpected natures of the thermal JCM itself is
our motivation for this work.

In the current paper, we discuss the following two facts:
\begin{itemize}
\item
If we adjust the time displacement $\Delta t$ properly,
figures consisting of plotted dots at the constant interval $\Delta t$
acquire scale invariance under replacement of $\Delta t$
by $s\Delta t$,
where $s(>1)$ is an arbitrary real but not transcendental number.
This fact is derived by making use of the Weyl criterion,
which gives a necessary and sufficient condition for
a sequence of real numbers to be uniformly distributed modulo unity.
\item
We can compute values of the time variable $t$,
which let $|S_{z}(t)|$
(the absolute value of the $z$-component of the Bloch vector)
be very small,
with the Diophantine approximation
(a rational approximation of an irrational number).
\end{itemize}

In Ref.~\cite{Azuma2008}, Azuma examines a histogram of
$\{S_{z}(n\Delta t):n=0,1,...,N\}$,
where $\mbox{\boldmath $S$}(t)$ is the Bloch vector of the thermal JCM.
Plotting the variance of the histogram of samples $\{S_{z}(n\Delta t)\}$
against the inverse of the temperature $\beta[=1/(k_{\mbox{\scriptsize B}}T)]$,
Azuma finds a scaling property.
In the current paper, we also take a constant interval $\Delta t$ for the time variable $t$,
and this prescription takes an important role.
Thus, the present paper is a sequel of Ref.~\cite{Azuma2008}.

This paper is organized as follows:
In Sec.~\ref{section-review-JCM},
we give a brief review of the thermal JCM.
In Sec.~\ref{section-trajectory-Bloch-vector},
we examine trajectories of the Bloch vector numerically.
In Sec.~\ref{section-quasiperiodicity},
we explain the quasiperiodicity, which we can observe in the trajectories of the Bloch vector.
In Sec.~\ref{section-scale-invariance},
we discuss the scale invariance of the figures generated as discrete plots of the Bloch vector.
In Secs.~\ref{section-L4-zero-points-numerical-approach} and \ref{section-perturbative-approach},
we investigate a graph of $S_{x}(t)$ versus the inverse of the temperature $\beta$
for the time $t$ such that $|S_{z}(t)|\ll 1$
by the numerical experiments and the perturbative evaluation, respectively.
[$\mbox{\boldmath $S$}(t)$ stands for the Bloch vector.]
In Sec.~\ref{section-discussions}, we give brief discussions.
In Appendix~\ref{section-appendix-A},
we consider physical transient spectra of the atom in the cavity.

\section{\label{section-review-JCM}
A brief review of the thermal JCM}
In this section, we give a brief review of the thermal JCM.
To describe its dynamics,
we use the notation of Ref.~\cite{Azuma2008}.

The Hamiltonian of the JCM is expressed in the form,
\begin{equation}
H=\frac{\hbar}{2}\omega_{0}\sigma_{z}+\hbar\omega a^{\dagger}a
+\hbar g(\sigma_{+}a+\sigma_{-}a^{\dagger}),
\label{JCM-Hamiltonian}
\end{equation}
\begin{equation}
\sigma_{\pm}=\frac{1}{2}(\sigma_{x}\pm i\sigma_{y}),
\label{definition-sigma+-}
\end{equation}
\begin{equation}
\sigma_{x}=
\left(
\begin{array}{cc}
0 & 1 \\
1 & 0
\end{array}
\right),
\quad
\sigma_{y}=
\left(
\begin{array}{cc}
0 & -i \\
i & 0
\end{array}
\right),
\quad
\sigma_{z}=
\left(
\begin{array}{cc}
1 & 0 \\
0 & -1
\end{array}
\right),
\label{definition-Pauli-matrices}
\end{equation}
\begin{equation}
[a,a^{\dagger}]=1,
\quad\quad
[a,a]=[a^{\dagger},a^{\dagger}]=0,
\label{commutation-relations-creation-annihilation-operators-photons}
\end{equation}
where $\sigma_{x}$, $\sigma_{y}$ and $\sigma_{z}$ given by
Eqs.~(\ref{definition-sigma+-}) and (\ref{definition-Pauli-matrices}) are the Pauli matrices
acting on atomic state vectors,
and $a^{\dagger}$ and $a$ given by
Eq.~(\ref{commutation-relations-creation-annihilation-operators-photons})
are the photon creation and annihilation operators, respectively.
We write the state of the two-level atom as a two-component column vector.
We describe the state of the cavity field as a superposition of number states of photons.

Let us divide the JCM Hamiltonian defined in Eq.~(\ref{JCM-Hamiltonian})
into two parts as follows:
\begin{eqnarray}
H&=&\hbar(C_{1}+C_{2}), \nonumber \\
C_{1}&=&\omega(\frac{1}{2}\sigma_{z}+a^{\dagger}a), \nonumber \\
C_{2}&=&g(\sigma_{+}a+\sigma_{-}a^{\dagger})-\frac{\Delta\omega}{2}\sigma_{z},
\end{eqnarray}
where $\Delta\omega=\omega-\omega_{0}$.
Then, we obtain the following relation:
\begin{equation}
[C_{1},C_{2}]=0.
\label{commutation-relation-C1-C2}
\end{equation}

Because we can diagonalize the Hermitian operator $C_{1}$ at ease,
we take the following interaction picture for describing the state of both the atom and the field.
First, we write a state vector of the whole system in the Schr{\"o}dinger and interaction
pictures as $|\psi_{\mbox{\scriptsize S}}(t)\rangle$ and $|\psi_{\mbox{\scriptsize I}}(t)\rangle$,
respectively.
Second, assuming $|\psi_{\mbox{\scriptsize I}}(0)\rangle=|\psi_{\mbox{\scriptsize S}}(0)\rangle$,
we define $|\psi_{\mbox{\scriptsize I}}(t)\rangle$ as follows:
\begin{equation}
|\psi_{\mbox{\scriptsize I}}(t)\rangle=\exp(iC_{1}t)|\psi_{\mbox{\scriptsize S}}(t)\rangle.
\label{definition-interaction-rep}
\end{equation}
Thus, because of Eq.~(\ref{commutation-relation-C1-C2}),
the time evolution of $|\psi_{\mbox{\scriptsize I}}(t)\rangle$ is given by
\begin{equation}
|\psi_{\mbox{\scriptsize I}}(t)\rangle=U(t)|\psi(0)\rangle,
\label{evolution-interaction-rep}
\end{equation}
where
\begin{equation}
U(t)=\exp(-iC_{2}t).
\label{evolution-unitary-operator}
\end{equation}

Here, we give an explicit form of $U(t)$ as follows:
\begin{eqnarray}
U(t)
&=&
\exp[-it
\left(
\begin{array}{cc}
-\Delta\omega/2 & ga \\
ga^{\dagger} & \Delta\omega/2
\end{array}
\right)
] \nonumber \\
&=&
\sum_{n=0}^{\infty}\frac{(-1)^{n}t^{2n}}{(2n)!}
\left(
\begin{array}{cc}
(D+g^{2})^{n} & 0 \\
0 & D^{n}
\end{array}
\right) \nonumber \\
&&
+
\sum_{n=0}^{\infty}\frac{(-i)(-1)^{n}t^{2n+1}}{(2n+1)!}
\left(
\begin{array}{cc}
-(\Delta\omega/2)(D+g^{2})^{n} & gaD^{n} \\
ga^{\dagger}(D+g^{2})^{n} & (\Delta\omega/2)D^{n}
\end{array}
\right) \nonumber \\
&=&
\left(
\begin{array}{cc}
u_{00} & u_{01} \\
u_{10} & u_{11}
\end{array}
\right), \label{explicit-time-evolution-op}
\end{eqnarray}
\begin{equation}
D=(\frac{\Delta\omega}{2})^{2}+g^{2}a^{\dagger}a,
\label{definition-op-D}
\end{equation}
\begin{eqnarray}
u_{00}&=&
\cos(t\sqrt{D+g^{2}})
+\frac{i}{2}\Delta\omega\frac{\sin(t\sqrt{D+g^{2}})}{\sqrt{D+g^{2}}}, \nonumber \\
u_{01}&=&
-iga\frac{\sin(t\sqrt{D})}{\sqrt{D}}, \nonumber \\
u_{10}&=&
-iga^{\dagger}\frac{\sin(t\sqrt{D+g^{2}})}{\sqrt{D+g^{2}}}, \nonumber \\
u_{11}&=&
\cos(t\sqrt{D})
-\frac{i}{2}\Delta\omega\frac{\sin(t\sqrt{D})}{\sqrt{D}}.
\label{elements-time-evolution-op}
\end{eqnarray}
We put the whole system into the following initial states:
\begin{equation}
\rho_{\mbox{\scriptsize AP}}(0)
=\rho_{\mbox{\scriptsize A}}(0)\otimes\rho_{\mbox{\scriptsize P}},
\label{whole-system-initial-state}
\end{equation}
\begin{equation}
\rho_{\mbox{\scriptsize A}}(0)
=\sum_{i,j\in\{0,1\}}\rho_{\mbox{\scriptsize A},ij}(0)
|i\rangle_{\mbox{\scriptsize A}}{}_{\mbox{\scriptsize A}}\langle j|,
\label{atom-initial-state}
\end{equation}
\begin{eqnarray}
\rho_{\mbox{\scriptsize P}}
&=&
\frac{\exp(-\beta\hbar\omega a^{\dagger}a)}{\mbox{Tr}\exp(-\beta\hbar\omega a^{\dagger}a)} \nonumber \\
&=&
(1-e^{-\beta\hbar\omega})\exp(-\beta\hbar\omega a^{\dagger}a),
\label{field-initial-state}
\end{eqnarray}
\begin{equation}
|0\rangle_{\mbox{\scriptsize A}}
=
\left(
\begin{array}{c}
1 \\
0
\end{array}
\right),
\quad
|1\rangle_{\mbox{\scriptsize A}}
=
\left(
\begin{array}{c}
0 \\
1
\end{array}
\right),
\label{atom-basis-vector}
\end{equation}
where $\{\rho_{\mbox{\scriptsize A},ij}\}$ represent
an arbitrary $2\times 2$ Hermitian matrix
with nonnegative eigenvalues and trace unity.
The indices A and P stand for the atom and the photon, respectively.
The density operator $\rho_{\mbox{\scriptsize P}}$ given by Eq.~(\ref{field-initial-state})
represents the thermal equilibrium state weighted by the Bose-Einstein distribution
for the inverse of the temperature $\beta[=1/(k_{\mbox{\scriptsize B}}T)]$.

After these preparations,
the time evolution of the atomic density operator is given by
\begin{equation}
\rho_{\mbox{\scriptsize A}}(t)
=\sum_{i,j\in\{0,1\}}\rho_{\mbox{\scriptsize A},ij}(t)
|i\rangle_{\mbox{\scriptsize A}}{}_{\mbox{\scriptsize A}}\langle j|,
\label{atom-evolution1}
\end{equation}
\begin{eqnarray}
\rho_{\mbox{\scriptsize A},ij}(t)
&=&\sum_{k,l\in\{0,1\}}\rho_{\mbox{\scriptsize A},kl}(0)
A_{kl,ij}(t)
\quad
\mbox{for $i,j\in\{0,1\}$},
\label{atom-evolution2}
\end{eqnarray}
\begin{equation}
A_{kl,ij}(t)
={}_{\mbox{\scriptsize A}}\langle i|\mbox{Tr}_{\mbox{\scriptsize P}}
[U(t)(|k\rangle_{\mbox{\scriptsize A}}{}_{\mbox{\scriptsize A}}\langle l|\otimes \rho_{\mbox{\scriptsize P}})
U^{\dagger}(t)]|j\rangle_{\mbox{\scriptsize A}}.
\label{atom-evolution3}
\end{equation}
Here, we pay attention to the following facts.
Clearly, relations
$\rho_{\mbox{\scriptsize A},10}(t)=\rho_{\mbox{\scriptsize A},01}(t)^{*}$
and
$\rho_{\mbox{\scriptsize A},11}(t)=1-\rho_{\mbox{\scriptsize A},00}(t)$
hold.
Thus, we need to estimate only two components,
$\rho_{\mbox{\scriptsize A},00}(t)$
and
$\rho_{\mbox{\scriptsize A},01}(t)$,
so that
\begin{eqnarray}
A_{00,00}(t)&=&
(1-e^{-\beta\hbar\omega}) \nonumber \\
&&\quad
\times
\sum_{n=0}^{\infty}
\frac{(\Delta\omega/2)^{2}+g^{2}(n+1)\cos^{2}(t\sqrt{\tilde{D}(n+1)})}{\tilde{D}(n+1)}
e^{-n\beta\hbar\omega}, \nonumber \\
A_{11,00}(t)&=&
(1-e^{-\beta\hbar\omega})
\sum_{n=1}^{\infty}
g^{2}n
\frac{\sin^{2}(t\sqrt{\tilde{D}(n)})}{\tilde{D}(n)}
e^{-n\beta\hbar\omega}, \nonumber \\
A_{01,01}(t)&=&
(1-e^{-\beta\hbar\omega})
\sum_{n=0}^{\infty}
[\cos(t\sqrt{\tilde{D}(n+1)})
+\frac{i}{2}\Delta\omega\frac{\sin(t\sqrt{\tilde{D}(n+1)})}{\sqrt{\tilde{D}(n+1)}}] \nonumber \\
&&\times
[\cos(t\sqrt{\tilde{D}(n)})
+\frac{i}{2}\Delta\omega\frac{\sin(t\sqrt{\tilde{D}(n)})}{\sqrt{\tilde{D}(n)}}]e^{-n\beta\hbar\omega},
\label{A-elements-explicit-form1}
\end{eqnarray}
\begin{equation}
A_{01,00}(t)=A_{10,00}(t)=A_{00,01}(t)=A_{10,01}(t)=A_{11,01}(t)=0,
\label{A-elements-explicit-form2}
\end{equation}
\begin{equation}
\tilde{D}(n)=(\frac{\Delta\omega}{2})^{2}+g^{2}n.
\label{definition-tilde-D}
\end{equation}

Then, we introduce the Bloch vector $\mbox{\boldmath $S$}(t)=(S_{x}(t),S_{y}(t),S_{z}(t))$,
which provides us with a visual description of the dynamics of the atomic state in a convenient way,
\begin{equation}
\rho_{\mbox{\scriptsize A}}(t)
=
\frac{1}{2}[\mbox{\boldmath $I$}+\mbox{\boldmath $S$}(t)\cdot\mbox{\boldmath $\sigma$}],
\label{definition-Bloch-vector}
\end{equation}
\begin{equation}
\mbox{\boldmath $S$}(t)
=
\left(
\begin{array}{ccc}
L_{\Delta\omega}^{(1)}(t) & L_{\Delta\omega}^{(2)}(t) & 0 \\
-L_{\Delta\omega}^{(2)}(t) & L_{\Delta\omega}^{(1)}(t) & 0 \\
0 & 0 & L_{\Delta\omega}^{(3)}(t)
\end{array}
\right)
\mbox{\boldmath $S$}(0)
+
\left(
\begin{array}{c}
0 \\
0 \\
L_{\Delta\omega}^{(4)}(t)
\end{array}
\right),
\label{equation-time-evolution-Bloch-vector-general}
\end{equation}
\begin{eqnarray}
L_{\Delta\omega}^{(1)}(t)&=&\mbox{Re}[A_{01,01}(t)], \nonumber \\
L_{\Delta\omega}^{(2)}(t)&=&\mbox{Im}[A_{01,01}(t)], \nonumber \\
L_{\Delta\omega}^{(3)}(t)&=&A_{00,00}(t)-A_{11,00}(t), \nonumber \\
L_{\Delta\omega}^{(4)}(t)&=&A_{00,00}(t)+A_{11,00}(t)-1.
\label{elements-equation-evolution-Bloch-vector-general}
\end{eqnarray}

\section{\label{section-trajectory-Bloch-vector}
Trajectories of the Bloch vector}
In this section, we examine the time evolution of the Bloch vector
given by Eqs.~(\ref{equation-time-evolution-Bloch-vector-general}) and
(\ref{elements-equation-evolution-Bloch-vector-general})
numerically.
From now on, for simplicity,
we assume $\Delta\omega=0$.
Then, we can rewrite the equations that represent the time evolution of the Bloch vector as follows:
\begin{equation}
\mbox{\boldmath $S$}(t)
=
\left(
\begin{array}{ccc}
L_{1}(t) & 0 & 0 \\
0 & L_{1}(t) & 0 \\
0 & 0 & L_{3}(t)
\end{array}
\right)
\mbox{\boldmath $S$}(0)
+
\left(
\begin{array}{c}
0\\
0\\
L_{4}(t)
\end{array}
\right),
\label{evolution-Bloch-vector-resonant}
\end{equation}
\begin{eqnarray}
L_{1}(t)&=&
(1-e^{-\beta})
\sum_{n=0}^{\infty}
\cos(\sqrt{n+1}t)\cos(\sqrt{n}t)e^{-n\beta}, \nonumber \\
L_{3}(t)&=&
\frac{1}{2}(1-e^{-\beta})
+\frac{e^{2\beta}-1}{2e^{\beta}}
\sum_{n=1}^{\infty}
\cos(2\sqrt{n}t)e^{-n\beta}, \nonumber \\
L_{4}(t)&=&
-\frac{1}{2}(1-e^{-\beta})
+\frac{(e^{\beta}-1)^{2}}{2e^{\beta}}
\sum_{n=1}^{\infty}
\cos(2\sqrt{n}t)e^{-n\beta},
\label{definition-L1-L3-L4}
\end{eqnarray}
where we assume $\omega\neq 0$ and $g\neq 0$,
and we replace parameters $\beta\hbar\omega$ and $t|g|$
with $\beta$ and $t$, respectively.
They imply that the time $t$ is in units of $|g|^{-1}$ and
the inverse of the temperature $\beta$ is in units of $(\hbar\omega)^{-1}$.
As a result of these replacements,
the Bloch vector $\mbox{\boldmath $S$}(t)$ depends only on two dimensionless variables, $t$ and $\beta$.

Here, putting the initial state of the atom into
$(1/\sqrt{2})(|0\rangle_{\mbox{\scriptsize A}}+|1\rangle_{\mbox{\scriptsize A}})$,
we obtain the initial Bloch vector $\mbox{\boldmath $S$}(0)=(1,0,0)$
and its time evolution
\begin{equation}
\mbox{\boldmath $S$}(t)
=
\left(
\begin{array}{c}
L_{1}(t)\\
0\\
L_{4}(t)
\end{array}
\right).
\label{Bloch-xz-plane}
\end{equation}
Equation~(\ref{Bloch-xz-plane}) tells us that the Bloch vector $\mbox{\boldmath $S$}(t)$ always lies
on the $xz$-plane $\forall t(\geq 0)$,
so that it is convenient for tracing the trajectory of $\mbox{\boldmath $S$}(t)$
as time passes.
Thus, from now on, we only examine the case where the Bloch vector is given
by Eqs.~(\ref{definition-L1-L3-L4}) and (\ref{Bloch-xz-plane}).

\begin{figure}
\begin{center}
\includegraphics[scale=1.0]{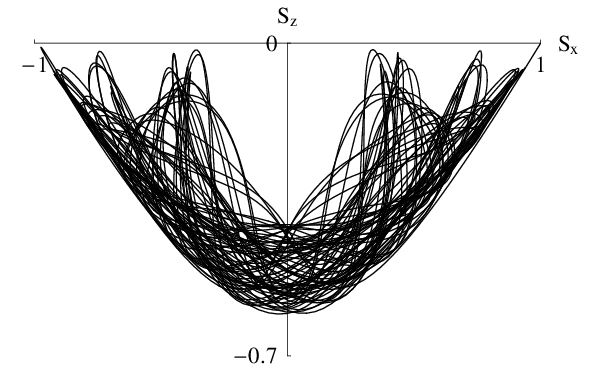}
\end{center}
\caption{A trajectory of $\mbox{\boldmath $S$}(t)$ given
by Eqs.~(\ref{definition-L1-L3-L4}) and (\ref{Bloch-xz-plane})
for $0\leq t \leq 250$ and $\beta=1.0$.}
\label{Figure01}
\end{figure}

Figure~\ref{Figure01} shows a trajectory of the Bloch vector $\mbox{\boldmath $S$}(t)$
given by Eqs.~(\ref{definition-L1-L3-L4}) and (\ref{Bloch-xz-plane})
for $\beta=1.0$ and $0\leq t\leq 250$.
To obtain Fig.~\ref{Figure01}, we replace an infinite summation $\sum_{n=0}^{\infty}$
with a finite summation $\sum_{n=0}^{150}$
for an actual numerical calculation of $L_{1}(t)$ defined in Eq.~(\ref{definition-L1-L3-L4}).
For calculating $L_{4}(t)$ numerically,
we give a similar treatment.
Throughout this paper,
whenever we carry out numerical calculations of $L_{1}(t)$ and $L_{4}(t)$,
we always apply this approximation to them.
At the end of this section,
we argue numerical errors given rise to by this approximation.

Looking at Fig.~\ref{Figure01}, we feel that both a norm and a direction of $\mbox{\boldmath $S$}(t)$
change hard at random and it is in a state of disorder.
With a careful observation of the trajectory,
we notice that the figure drawn by $\mbox{\boldmath $S$}(t)$ is nearly symmetrical
with respect to the $z$-axis and the trajectory lies within a particular limited area
on the $xz$-plane.
However, it is difficult for us to find a regular form and order any more from Fig.~\ref{Figure01}.

As shown in Fig.~\ref{Figure01},
introducing the non-zero temperature and its thermal fluctuation,
we can observe that the behaviour of $\mbox{\boldmath $S$}(t)$ becomes in disorder.
However, if we look at the time evolution of $\mbox{\boldmath $S$}(t)$
from a new viewpoint, which is different from ones that we have been used to,
we can find novel structures hidden in the trajectory of $\mbox{\boldmath $S$}(t)$.
We show these secret structures in the following paragraphs.

\begin{figure}
\begin{center}
\includegraphics[scale=1.0]{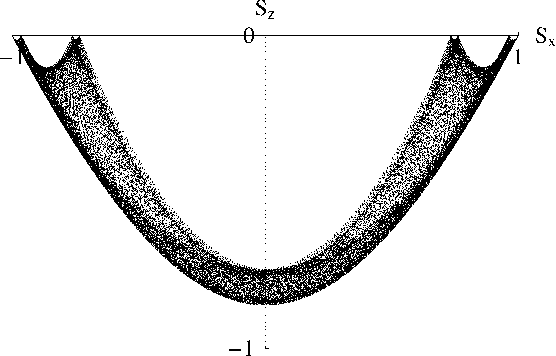}
\end{center}
\caption{A figure consisting of dots of $\mbox{\boldmath $S$}(t)$ given
by Eqs.~(\ref{definition-L1-L3-L4}) and (\ref{Bloch-xz-plane}).
We plot them at the constant time interval $\Delta t=3.5$ for $\beta=2.0$.
The number of dots is equal to $N=128\mbox{ }000$.
We give all points a diameter being $1/1000$ of the width of the whole graph.}
\label{Figure02}
\end{figure}

\begin{figure}
\begin{center}
\includegraphics[scale=1.0]{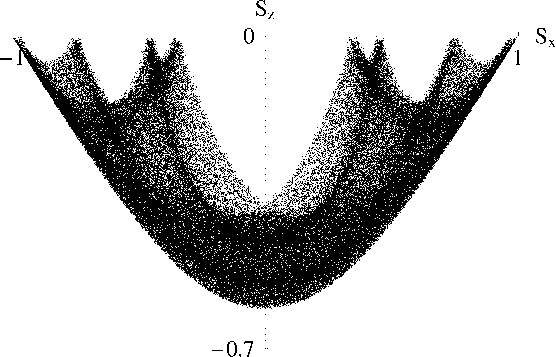}
\end{center}
\caption{A figure consisting of dots of $\mbox{\boldmath $S$}(t)$ given
by Eqs.~(\ref{definition-L1-L3-L4}) and (\ref{Bloch-xz-plane}).
We plot them at the constant time interval $\Delta t=3.5$ for $\beta=1.0$.
The number of dots is equal to $N=384\mbox{ }000$.
We give all points a diameter being $1/1000$ of the width of the whole graph.}
\label{Figure03}
\end{figure}

\begin{figure}
\begin{center}
\includegraphics[scale=1.0]{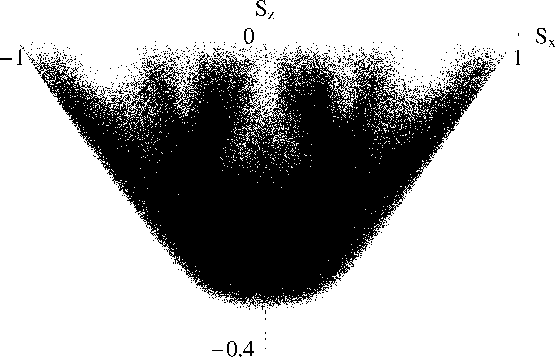}
\end{center}
\caption{A figure consisting of dots of $\mbox{\boldmath $S$}(t)$ given
by Eqs.~(\ref{definition-L1-L3-L4}) and (\ref{Bloch-xz-plane}).
We plot them at the constant time interval $\Delta t=3.5$ for $\beta=0.5$.
The number of dots is equal to $N=1\mbox{ }024\mbox{ }000$.
We give all points a diameter being $1/1000$ of the width of the whole graph.}
\label{Figure04}
\end{figure}

We plot trajectories of the Bloch vector
given by Eqs.~(\ref{definition-L1-L3-L4}) and (\ref{Bloch-xz-plane})
on the $xz$-plane as $\mbox{\boldmath $S$}(t)=(S_{x}(t),S_{z}(t))$
in Figs.~\ref{Figure02}, \ref{Figure03} and \ref{Figure04}.
In these figures,
we plot $\mbox{\boldmath $S$}(t)$ at a constant time interval,
so that the time variable takes discrete values as $t_{n}=n\Delta t$ for $n=0, 1, 2, ..., N$.
Moreover, we have the number of dots plotted $N$ as large as possible.
Turning our eyes on Figs.~\ref{Figure02}, \ref{Figure03} and \ref{Figure04},
we understand that the trajectories of the Bloch vector reveal
regular forms, distinct orders and novel structures,
that we have not been able to discover before.
 
\begin{figure}
\begin{center}
\includegraphics[scale=1.0]{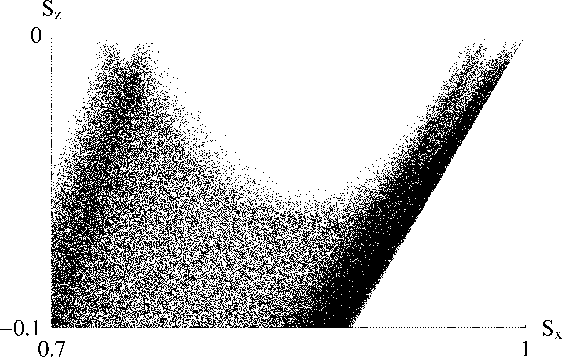}
\end{center}
\caption{A figure consisting of dots of $\mbox{\boldmath $S$}(t)$ given
by Eqs.~(\ref{definition-L1-L3-L4}) and (\ref{Bloch-xz-plane}).
We plot them at the constant time interval $\Delta t=3.5$ for $\beta=1.0$
and $N=12\mbox{ }800\mbox{ }000$.
The horizontal and vertical ranges are given by
$0.7\leq S_{x}\leq 1.0$ and $-0.1\leq S_{z}\leq 0.0$, respectively.
This figure corresponds to an enlargement of the upper right part of Fig.~\ref{Figure03}.
The number of dots plotted actually for this figure is equal to $274\mbox{ }692$.
We give all points a diameter being $1/1000$ of the width of the whole graph.}
\label{Figure05}
\end{figure}

\begin{figure}
\begin{center}
\includegraphics[scale=1.0]{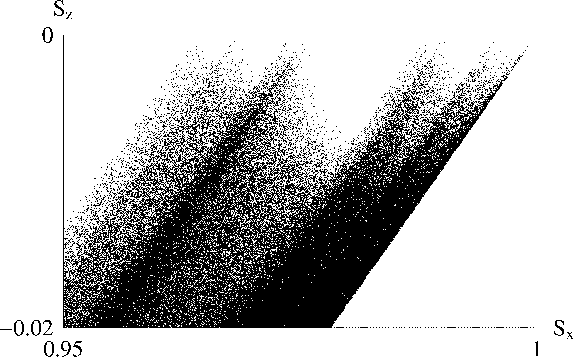}
\end{center}
\caption{A figure consisting of dots of $\mbox{\boldmath $S$}(t)$ given
by Eqs.~(\ref{definition-L1-L3-L4}) and (\ref{Bloch-xz-plane}).
We plot them at the constant time interval $\Delta t=3.5$ for $\beta=1.0$
and $N=1\mbox{ }024\mbox{ }000\mbox{ }000$.
The horizontal and vertical ranges are given by
$0.95\leq S_{x}\leq 1.0$ and $-0.02\leq S_{z}\leq 0.0$, respectively.
This figure corresponds to an enlargement of the upper right part of Fig.~\ref{Figure05}.
The number of dots plotted actually for this figure is equal to $507\mbox{ }596$.
We give all points a diameter being $1/1000$ of the width of the whole graph.}
\label{Figure06}
\end{figure}

\begin{figure}
\begin{center}
\includegraphics[scale=1.0]{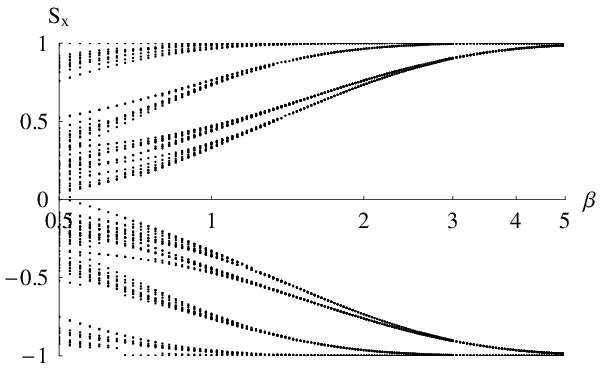}
\end{center}
\caption{Taking a certain positive number being small enough as $0<\epsilon\ll 1$,
we plot $(\beta,S_{x}(t_{n}))$ for discrete time variable $t_{n}=n\Delta t$ for $\Delta t=3.5$
on condition that
$|S_{z}(t_{n})|\leq\epsilon$ holds.
The horizontal axis represents $\beta$ for the range of $0.5\leq\beta\leq 5.0$
with the logarithmic scale.
We plot the points at constant intervals,
$\Delta\beta=0.025$ for $0.5\leq\beta\leq 3$
and
$\Delta\beta=0.05$ for $3\leq\beta\leq 5$.
We give them a diameter being $4/1000$ of the width of the whole graph.
(We explain how to draw this graph precisely
in Sec.~\ref{section-L4-zero-points-numerical-approach}.)}
\label{Figure07}
\end{figure}

Magnifying the upper right part of Fig.~\ref{Figure03},
we obtain Fig.~\ref{Figure05}.
And magnifying the upper right part of Fig.~\ref{Figure05},
we obtain Fig.~\ref{Figure06}.
Looking at Figs.~\ref{Figure02}, \ref{Figure03}, \ref{Figure04}, \ref{Figure05} and \ref{Figure06}
carefully,
we notice that points of $|S_{z}|\ll 1$
increase in number as $\beta$ becomes smaller, that is, the temperature becomes larger.
Moreover, we observe the following facts:
On the one hand, points of $|S_{z}|\ll 1$ for $\beta\gg 1$
satisfy $|S_{x}|\simeq 1$.
On the other hand,
points of $|S_{z}|\ll 1$ for $0<\beta\ll 1$ are distributed widely in the range of $-1\leq S_{x}\leq 1$.
In Fig.~\ref{Figure07}, we examine this tendency of the Bloch vector $\mbox{\boldmath $S$}(t)$
numerically.

How to draw a graph of Fig.~\ref{Figure07} is as follows:
First, we consider a set of samples,
\begin{equation}
\{S_{z}(t_{n}):n\in\{0,1,2,...,N\}\},
\end{equation}
where $t_{0}, t_{1}, ..., t_{N}$ are discrete values of the time variable generated
with the constant interval
$\Delta t$ as mentioned before.
Second, we take a certain positive number being small enough,
$0<\epsilon\ll 1$.
Third, we select values of $t_{n}$,
each of which satisfies $|S_{z}(t_{n})|\leq\epsilon$.
Because of Eqs.~(\ref{definition-L1-L3-L4}) and (\ref{Bloch-xz-plane}),
we can regard $S_{x}(=L_{1})$ as a function depending only on two variables,
$t_{n}$ and $\beta$.
Fourth, we plot $(\beta,S_{x}(t_{n}))$ such that $|S_{z}(t_{n})|\leq\epsilon$
and obtain the graph of Fig.~\ref{Figure07}.
(In Sec.~\ref{section-L4-zero-points-numerical-approach},
we explain details for drawing Fig.~\ref{Figure07},
for example, how to choose $\epsilon$ and $N$ properly
for given $\beta$ such that $0.5\leq \beta\leq 5.0$,
with giving a concrete example.)
Figure~\ref{Figure07} shows dependence of $S_{x}$ on $\beta$ on condition that $|S_{z}|\ll 1$ holds.
In the graph of Fig.~\ref{Figure07},
we can observe successive splits leading to different branches.
It gives us an impression that the Bloch vector $\mbox{\boldmath $S$}(t)$
hides further secret orders and structures inside itself.

Closing this section, let us argue the numerical precisions of $L_{1}(t)$ and $L_{4}(t)$.
As mentioned before,
whenever we carry out the numerical calculations of $L_{1}(t)$ and $L_{4}(t)$
defined in Eq.~(\ref{definition-L1-L3-L4}),
we replace the infinite summation $\sum_{n=0}^{\infty}$ with the finite summation $\sum_{n=0}^{150}$,
which contains the first $151$ terms.
Here, we estimate numerical errors caused by this treatment.
For example, we can evaluate the upper bound of numerical errors for $L_{1}(t)$ as
\begin{eqnarray}
E[L_{1}]
&\leq&
(1-e^{-\beta})|\sum_{n=151}^{\infty}\cos(\sqrt{n+1}t)\cos(\sqrt{n}t)e^{-n\beta}| \nonumber \\
&\leq&
(1-e^{-\beta})\sum_{n=151}^{\infty}e^{-n\beta} \nonumber \\
&=&
(1-e^{-\beta})e^{-151\beta}\sum_{n=0}^{\infty}e^{-n\beta} \nonumber \\
&=&
e^{-151\beta}.
\label{upper-bound-erroe-L1-0}
\end{eqnarray}
Thus, if we assume $0.5\leq\beta\leq 5.0$, we obtain
\begin{equation}
E[L_{1}]\leq 1.63\times 10^{-33}.
\label{upper-bound-erroe-L1-1}
\end{equation}
Therefore, our calculation of $L_{1}(t)$ serves a precision of $33$ significant decimal digits to us.
Similar things hold for $L_{4}(t)$.

To obtain numerical data for drawing graphs of
Figs.~\ref{Figure01}, \ref{Figure02}, \ref{Figure03}, \ref{Figure04}, \ref{Figure05}, \ref{Figure06}
and \ref{Figure07},
we utilize the Fortran compiler,
which can manipulate the data type of the real quadruple precision.
This implies that the compiler sorts out numerical calculations
with a precision of about $33$ significant decimal digits,
so that it is fit to suppress numerical errors
as Eqs.~(\ref{upper-bound-erroe-L1-0}) and (\ref{upper-bound-erroe-L1-1}).

To obtain numerical data for drawing graphs of
Figs.~\ref{Figure11}, \ref{Figure12}, \ref{Figure13}, \ref{Figure14}
and \ref{Figure15},
and to carry out calculations required in Sec.~\ref{section-perturbative-approach},
we utilize a high-performance computer algebra system, the Mathematica.
The Mathematica has an option allowing us to specify any precision we want,
so that we let it execute numerical calculations
with $50$ significant decimal digits for keeping extra enough amount of digits.
In Sec.~\ref{section-perturbative-approach}, we often have to manipulate integers,
whose number of decimal digits is equal to $15$,
in computations.
Under these circumstances,
we need $15$ digits for integers and $33$ digits for the decimal representation of the floating point,
so that we have to prepare $48$ digits in total for computations over real numbers.
Thus, utilizing the Mathematica with the $50$ significant decimal digits,
we can keep away from danger caused by the roundoff and truncation errors.
Therefore, we can obtain numerical values of $L_{1}(t)$ and $L_{4}(t)$ with $33$ significant decimal digits
at least,
and we can obtain stable results of the numerical calculations.

\section{\label{section-quasiperiodicity}
Quasiperiodicity in the dynamics of the Bloch vector}
In this section, we discuss the quasiperiodicity of the Bloch vector,
which is given by Eqs.~(\ref{evolution-Bloch-vector-resonant}) and (\ref{definition-L1-L3-L4}).
Here, we examine the behaviour of the Bloch vector under the low temperature limit.
Assuming that the temperature is low enough as $\beta=1/(k_{\mbox{\scriptsize B}}T)\gg 1$
and neglecting the terms of order $O((e^{-\beta})^{2})$,
we obtain approximations of $L_{i}(t)$ for $i=1, 3, 4$ as follows:
\begin{eqnarray}
L_{1}(t)
&\simeq&
(1-e^{-\beta})\cos t+e^{-\beta}\cos(\sqrt{2}t)\cos t,
\label{L1t-approximation-0} \\
L_{3}(t)
&\simeq&
\frac{1}{2}(1-e^{-\beta})+\frac{1}{2}\cos(2t),
\label{L3t-approximation-0} \\
L_{4}(t)
&\simeq&
-\frac{1}{2}(1-e^{-\beta})+\frac{1}{2}(1-2e^{-\beta})\cos(2t).
\label{L4t-approximation-0}
\end{eqnarray}

The above approximations of $L_{3}(t)$ and $L_{4}(t)$ are periodic functions of $t$ with period $\pi$.
In contrast, the function $L_{1}(t)$
given by Eq.~(\ref{L1t-approximation-0}) is made of $\cos t$ and $\cos(\sqrt{2}t)$,
that is,
a complex structure of trigonometric functions with periods $2\pi$ and $\sqrt{2}\pi$,
so that its behaviour is unpredictable.
Thus, from now on, we concentrate on analysing the function $L_{1}(t)$.
Equation (\ref{L1t-approximation-0}), the approximation of $L_{1}(t)$,
has two angular frequencies,
$\omega_{1}=1$ and $\omega_{2}=\sqrt{2}$,
so that the ratio of $\omega_{2}$ to $\omega_{1}$ is irrational as $\omega_{2}/\omega_{1}=\sqrt{2}$.
We call $m$ angular frequencies $\omega_{1}, \omega_{2}, ..., \omega_{m}$ are incommensurate
if no one of the angular frequencies $\omega_{i}$ can be expressed as a linear combination of the others
using coefficients that are rational numbers.
And if the system has incommensurate angular frequencies,
we call it quasiperiodic.

The notion of the quasiperiodicity means an intermediate state between periodic and chaotic ones
\cite{Ott1993,Ruelle1984,Goldstein2002,Ostlund1983,Glazier1988,Arnold2006}.
To understand the
quasiperiodicity of Eq.~(\ref{L1t-approximation-0})
giving the approximation of $L_{1}(t)$,
we examine a dynamical system that has two incommensurate angular frequencies,
$\omega_{1}$ and $\omega_{2}$.
In the case of two-angular-frequency quasiperiodic motion,
we can describe a dynamical variable $f(t)$ as a function of two independent variables,
$G(t_{1},t_{2})$,
for example,
\begin{equation}
f(t)
=
L_{1}(t)
=
(1-e^{-\beta})\cos t+e^{-\beta}\cos(\sqrt{2}t)\cos t,
\label{f-t-L1-first-order-0}
\end{equation}
\begin{equation}
G(t_{1},t_{2})
=
(1-e^{-\beta})\cos t_{1}+e^{-\beta}\cos(\sqrt{2}t_{2})\cos t_{1},
\label{definition-G-t1-t2-0}
\end{equation}
\begin{equation}
G(t_{1}+2\pi,t_{2})
=
G(t_{1},t_{2}+\sqrt{2}\pi)
=
G(t_{1},t_{2}),
\label{periodicity-G-t1-t2-0}
\end{equation}
\begin{equation}
f(t)
=
\left.
G(t_{1},t_{2})
\right|_{t_{1}=t_{2}=t}.
\label{quasiperiodicity-function-f-0}
\end{equation}
Letting $\omega_{1}=1$ and $\omega_{2}=\sqrt{2}$,
the equation of the form,
\begin{equation}
m_{1}\omega_{1}+m_{2}\omega_{2}=0,
\end{equation}
does not hold for arbitrary integers,
$m_{1}$ and $m_{2}$,
except for $m_{1}=m_{2}=0$,
so that $\omega_{1}$ and $\omega_{2}$ are incommensurate.
Thus, the dynamics of the system is specified by two independent angles $\theta_{1}$ and $\theta_{2}$
as
\begin{equation}
G(\frac{\theta_{1}}{\omega_{1}},\frac{\theta_{2}}{\omega_{2}}),
\end{equation}
where $0\leq\theta_{i}<2\pi$ for $i=1,2$.
This implies that the orbit of the motion lies on the torus embedded in the phase space.
From Eq.~(\ref{quasiperiodicity-function-f-0}),
we understand that $f(t)$ is the quantity obtained
along the orbit of $\theta_{1}=\omega_{1}t$ and $\theta_{2}=\omega_{2}t$
being realized on the torus in the phase space.
In the following, we consider these matters in a general manner.

Let us consider an arbitrary Hamiltonian
$H(q_{1},q_{2},p_{1},p_{2},t)$,
whose number of degrees of freedom is equal to two.
Because the system evolves according to this Hamiltonian,
the trajectory that the system follows is described with Hamilton's equations,
\begin{eqnarray}
\dot{q}_{i}
&=&
\frac{\partial H(\{q_{j}\},\{p_{j}\},t)}{\partial p_{i}}, \nonumber \\
\dot{p}_{i}
&=&
-
\frac{\partial H(\{q_{j}\},\{p_{j}\},t)}{\partial q_{i}}
\quad\quad
\mbox{for $i=1,2$}.
\end{eqnarray}

The phase space of the system is four-dimensional and its co-ordinate system is given by
$(q_{1}, q_{2}, p_{1}, p_{2})$.
Then, we remember the Poincar{\`e}-Cartan integral invariant and Liouville's theorem,
which are fundamental results
in the field of the analytical mechanics
\cite{Goldstein2002,Arnold2006,Jordan2004}.
They guarantee that the density of the system points in an infinitesimal volume element
is preserved
while they are travelling with time according to the canonical equations of motion.
In other words, the volume of a region of the phase space is invariant
under evolution with Hamilton's equations of the motion in time.

Now, we assume that the motion of the system is bounded in the four-dimensional phase space.
Furthermore, we suppose the Hamiltonian system to be integrable.
Thus, the system has two functions
$I_{i}(\{q_{j}\},\{p_{j}\})$ for $i=1,2$,
which are in involution and independent.
Moreover, these functions are time-independent and preserved.
Hence, they satisfy the following commutation relations:
\begin{equation}
[I_{1},I_{2}]=0,
\quad\quad
[I_{i},H]=0
\quad
\mbox{for $i=1,2$}.
\label{definition-integrals-of-motion-0}
\end{equation}
In the above commutation relations,
the bilinear operation $[u,v]$ denotes the Poisson bracket,
so that $I_{i}$ for $i=1,2$ are constants of the motion.

Here, let us define $I_{i}$ for $i=1,2$ at initial time ($t=0$) as follows:
\begin{equation}
I_{i}(0)=I_{i}(\{q_{j}(0)\},\{p_{j}(0)\})=\mbox{constant}
\quad
\mbox{for $i=1,2$}.
\end{equation}
This implies that the motion of the system is confined to a two-dimensional surface
embedded in the four-dimensional phase space $(q_{1}, q_{2}, p_{1}, p_{2})$.
In addition, the surface is determined by the two integrals of the motion,
$I_{1}(0)$ and $I_{2}(0)$,
and all possible states of the system have to lie on it.

From now on, we let ${\cal M}_{\mbox{\scriptsize \boldmath $I$}(0)}$ represent the above surface
specified with $I_{1}(0)$ and $I_{2}(0)$.
Because of the Poincar{\`e}-Cartan integral invariant and Liouville's theorem,
the motion of the integrable system bounded on the surface
${\cal M}_{\mbox{\scriptsize \boldmath $I$}(0)}$
has to be periodic.
This conclusion implies that the motion of the system is periodic along the trajectory
of the time evolution being determined by Hamilton's equations.

Moreover, we can extend this statement as follows.
At first, we assume that $I_{1}(0)$ and $I_{2}(0)$ are the integrals of the motion of the system.
Next, we regard these integrals as general angular momenta
$I_{i}(\{q_{j}\},\{p_{j}\})$ for $i=1,2$.
Let $\theta_{1}$ and $\theta_{2}$ be canonical angular variables of these angular momenta,
so that $(\theta_{1},I_{1})$ and $(\theta_{2},I_{2})$ form canonical co-ordinates.
Then, putting parameters $\theta_{i}$ along the loop as $\theta_{i}\in[0,2\pi)$ for $i=1,2$,
we can consider both variables of $\theta_{1}$ and $\theta_{2}$ to be periodic,
\begin{equation}
\theta_{i}=\omega_{i}t \pmod{2\pi}
\quad
\mbox{for $i=1,2$}.
\label{periodic-co-ordinates-0}
\end{equation}
We can regard Eq.~(\ref{periodic-co-ordinates-0}) as a definition of $\omega_{i}$ for $i=1,2$.

\begin{figure}
\begin{center}
\includegraphics[scale=1.0]{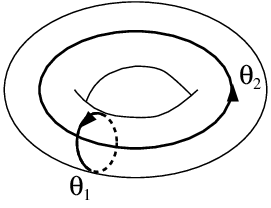}
\end{center}
\caption{We can consider the surface ${\cal M}_{\mbox{\scriptsize \boldmath $I$}(0)}$
to be topologically equivalent to a two-dimensional torus,
on which every point is specified with the co-ordinate system
$(\theta_{1}, \theta_{2})$.}
\label{Figure08}
\end{figure}

The reason why Eq.~(\ref{periodic-co-ordinates-0}) holds is as follows.
Here, we put a reference point (an origin) of the co-ordinate system $(\theta_{1},\theta_{2})$
at an initial point $\mu_{0}$ for $t=0$ on ${\cal M}_{\mbox{\scriptsize \boldmath $I$}(0)}$.
Moreover, we consider $(t_{1}I_{1}+t_{2}I_{2})$ to be a generator
of a canonical transformation.
Then, we can suppose $\mbox{\boldmath $t$}=(t_{1},t_{2})$ to be an arbitrary
two-dimensional real vector.
Hence, we understand that the two angles $\theta_{1}$ and $\theta_{2}$ are independent and periodic
in times $t_{1}$ and $t_{2}$, respectively.
Thus, we obtain $\theta_{i}=\omega_{i}t_{i}$ for $i=1,2$.
Then, reminding Eqs.~(\ref{f-t-L1-first-order-0}), (\ref{definition-G-t1-t2-0}),
(\ref{periodicity-G-t1-t2-0}) and (\ref{quasiperiodicity-function-f-0}),
we set $t_{1}=t_{2}=t$ and obtain Eq.~(\ref{periodic-co-ordinates-0}).
From the above discussion, we can consider that the surface ${\cal M}_{\mbox{\scriptsize \boldmath $I$}(0)}$
embedded in the four-dimensional phase space to be topologically equivalent to a two-dimensional torus,
$T^{2}=S^{1}\times S^{1}$,
as shown in Fig.~\ref{Figure08}.

\begin{figure}
\begin{center}
\includegraphics[scale=0.7]{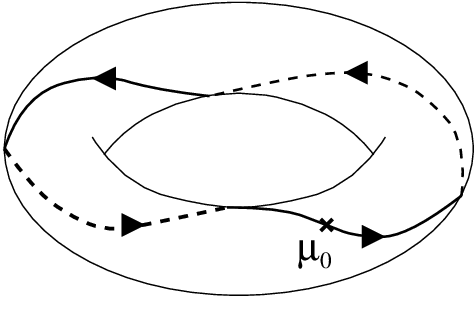}
\end{center}
\caption{In the case of $\omega_{2}/\omega_{1}=1/2$,
an orbit moves around the torus twice in the short direction $\theta_{1}$
and once in the long direction $\theta_{2}$ as time proceeds,
so that the orbit forms a finite closed path.}
\label{Figure09}
\end{figure}

Here, let us think about the special case $\omega_{2}/\omega_{1}=1/2$.
Then, an orbit winds around the torus twice in the short direction $\theta_{1}$ and simultaneously
once in the long direction $\theta_{2}$ as time progresses.
Thus, the orbit closes on itself after finite period of time as shown
in Fig.~\ref{Figure09}.
Hence, actually possible states of the system are confined
on the one-dimensional closed path lying on the torus.

\begin{figure}
\begin{center}
\includegraphics[scale=0.7]{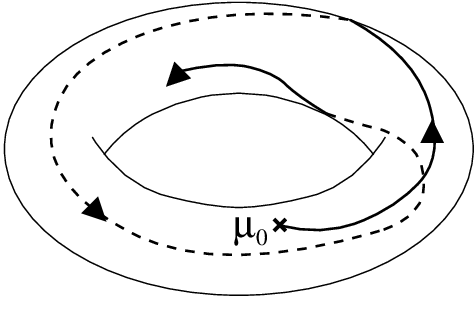}
\end{center}
\caption{In the case of $\omega_{2}/\omega_{1}=\sqrt{2}$,
an orbit never closes on itself in a finite time period.
The trajectory becomes dense on the torus as time proceeds to infinity.}
\label{Figure10}
\end{figure}

However, if we think about the case where $\omega_{2}/\omega_{1}$ is irrational,
for example,
$\omega_{2}/\omega_{1}=\sqrt{2}$,
the situation with respect to the orbit of the system becomes different drastically.
If $\omega_{2}/\omega_{1}$ is irrational,
the orbit never closes on itself as shown in Fig.~\ref{Figure10}.
As time goes to infinity,
the orbit on the torus will eventually come arbitrary close to every point on the toroidal surface.
Thus, the orbit fills up the torus in the phase space finally,
and the trajectory of the system becomes dense and uniformly distributed on the torus.

Then, we pay attention to the following fact.
As mentioned above, if $\omega_{2}/\omega_{1}$ is irrational,
the trajectory never closes on itself in the phase space.
However, taking the long-term limit,
the trajectory on the torus in the phase space will
eventually comes arbitrary close to every point on the toroidal surface.
Hence, if we take an arbitrary small positive number $\epsilon(>0)$,
there actually exists time $t$ such as
\begin{equation}
d(\mu_{0},\mu(t))<\epsilon,
\label{irrational-orbit-point-arraival-origin-0}
\end{equation}
where $\mu(t)$ denotes a state of the system on ${\cal M}_{\mbox{\scriptsize \boldmath $I$}(0)}$
at $t(>0)$
and $d(\mu_{0},\mu(t))$ represents a distance between $\mu_{0}$ (the origin) and $\mu(t)$
on ${\cal M}_{\mbox{\scriptsize \boldmath $I$}(0)}$.

The question is how to estimate time $t$,
which satisfies Eq.~(\ref{irrational-orbit-point-arraival-origin-0}).
Let us think about obtaining a rational approximate number of irrational $\omega_{2}/\omega_{1}$
by its continued fraction expression.
For example, in case of $\omega_{2}/\omega_{1}=\sqrt{2}$,
we can describe its continued fraction as
\begin{equation}
\sqrt{2}
=
1
+\frac{1}{\displaystyle 2
+\frac{1}{\displaystyle 2+ \frac{1}{\ddots}}}.
\label{continued-fraction-sqrt-2-0}
\end{equation}
By truncating the above expression with the first some terms,
we can obtain the rational approximate number $p/q$ with high precision such as
\begin{equation}
|\sqrt{2}-\frac{p}{q}|
<
\frac{1}{q^{2}}.
\end{equation}
(This fact is assured by the theorem of Roth
\cite{Coppel2009}.)

Thus, because of $\omega_{2}/\omega_{1}=\sqrt{2}\simeq p/q$,
we obtain $p\omega_{1}\simeq q\omega_{2}$.
This implies the following.
If the orbit winds around the torus $p$th times in the short direction $\theta_{1}$
and simultaneously $q$th times in the long direction $\theta_{2}$,
it arrives at a point which is very close to the initial point $\mu_{0}$.
We investigate this question in Sec.~\ref{section-perturbative-approach}.

In this section, so far, we suppose the low-temperature limit
and evaluate the contribution to $L_{1}(t)$ up to $O(e^{-\beta})$.
Here, we let the temperature become higher slightly and estimate the
contribution to $L_{1}(t)$ up to $O((e^{-\beta})^{2})$.
Then, we can regard $L_{1}(t)$ as a superposition of $\cos t$, $\cos(\sqrt{2}t)$ and $\cos(\sqrt{3}t)$,
and it causes quasiperiodicity of three incommensurate angular frequencies.
Moreover, if we make the temperature become much higher and estimate the contribution to $L_{1}(t)$ up to
$O((e^{-\beta})^{3})$,
$L_{1}(t)$ has four angular frequencies,
$\omega_{1}=1$, $\omega_{2}=\sqrt{2}$, $\omega_{3}=\sqrt{3}$ and $\omega_{4}=2$.
Then, $\omega_{4}/\omega_{1}$ is equal to two,
so that it is rational.
Let us think about the four-dimensional torus $T^{4}=S^{1}\times S^{1}\times S^{1}\times S^{1}$,
where the orbit of the motion lies, and its two-dimensional submanifold $(\theta_{1},\theta_{4})$.
Then, the orbit is periodic on the submanifold $(\theta_{1},\theta_{4})$.
So that, as the temperature becomes higher,
the motion of the Bloch vector causes a complicated trajectory,
which is a mixture of periodic and quasiperiodic orbits.

\section{\label{section-scale-invariance}
The scale invariance of the figures generated by the Bloch vector}
In this section, we show that the figures drawn
in Figs.~\ref{Figure02}, \ref{Figure03}, \ref{Figure04}, \ref{Figure05} and \ref{Figure06}
are invariant under a scale change in $\Delta t$.
We draw these figures by plotting the trajectories of the Bloch vector at the constant interval $\Delta t$.
To show the scale invariance of these discrete plots,
we use some properties of random sequences of real numbers,
which are uniformly distributed modulo $2\pi$.
In the former half of this section,
we give two considerations in preparation for explaining the scale invariance of the figures.
First, we discuss the physical meaning of plotting the trajectories of the Bloch vector
at the constant time interval $\Delta t$.
We examine differences between continuous and discrete plots.
Second, we discuss the random sequences that are uniformly distributed modulo $2\pi$.
After these preparations, in the latter half of this section,
we discuss the scale invariance of figures generated
by plotting the Bloch vector at the interval $\Delta t$.

First, we consider physical meaning of dividing time into a lattice.
With a discrete time slice $\Delta t$,
we have a chain of time co-ordinates $t_{n}=n\Delta t$ for $n=0, 1, 2, ..., N$.
Then, as shown in Figs.~\ref{Figure02}, \ref{Figure03}, \ref{Figure04}, \ref{Figure05} and \ref{Figure06},
we plot trajectories of the Bloch vector $\mbox{\boldmath $S$}(t)$ as
$\{\mbox{\boldmath $S$}(t_{n}):n\in\{0,1,2,...,N\}\}$.
Through this procedure,
we introduce a unit of time $\Delta t$.
From now on, we regard it as a scale of the time variable.
Indeed, assuming $\Delta t$ to be a real but not transcendental number and $\Delta t>\pi$,
the graphs of Figs.~\ref{Figure02}, \ref{Figure03}, \ref{Figure04}, \ref{Figure05} and \ref{Figure06}
are invariant under the following change of the scale,
\begin{equation}
\Delta t\to s\Delta t,
\label{rescaling-1}
\end{equation}
where $s$ is a real but not transcendental number and $s>1$.
We show this fact in the latter half of this section.

As an alternative plan to Eq.~(\ref{rescaling-1}),
we can consider a change of a scale of the continuous time variable,
\begin{equation}
t\to s t
\quad\quad
(s>0).
\label{rescaling-2}
\end{equation}
In this case, we handle the time variable $t$ as a continuous quantity.
Throughout the current paper, we adopt Eq.~(\ref{rescaling-1}) rather than Eq.~(\ref{rescaling-2}).
Here, we think about a difference between the change of the scale of Eq.~(\ref{rescaling-1}) and
that of Eq.~(\ref{rescaling-2}).
On the one hand,
if we choose Eq.~(\ref{rescaling-1}),
we have to neglect all events which relate to the quantity of the time being less than $\Delta t$.
On the other hand, if we select Eq.~(\ref{rescaling-2}),
we have to admit an infinitesimal time displacement for the system.
Thus, taking Eq.~(\ref{rescaling-1}),
we get rid of notions of infinitesimal time evolution.
In the latter half of this section,
we consider $\Delta t$ to be in the range between $\pi$ and $2\pi$, that is,
$\pi<\Delta t<2\pi$.
This treatment generates a random sequence
$t_{n}=n\Delta t \pmod{2\pi}$ for $n=0, 1, ..., N$ and
lets the graphs of discrete dots
in Figs.~\ref{Figure02}, \ref{Figure03}, \ref{Figure04}, \ref{Figure05} and \ref{Figure06}
be invariant under the change of the scale $\Delta t$.

Moreover, the above treatment produces an unexpected effect that removes a history of the sequence
$\{\mbox{\boldmath $S$}(t_{n}):t_{n}=n\Delta t,n\in\{0,1,2,...,N\}\}$
actually.
This is because $\Delta t$ is rather larger,
such as $\pi<\Delta t<2\pi$.
If we let the time displacement be small enough, the sequence of
$(\mbox{\boldmath $S$}(t_{0}), \mbox{\boldmath $S$}(t_{1}), \mbox{\boldmath $S$}(t_{2}),
...)$
approximates to the continuous trajectory of $\mbox{\boldmath $S$}(t)$ well.
However, if we take $\pi<\Delta t<2\pi$,
we cannot find neither trace nor history of continuous trajectory of $\mbox{\boldmath $S$}(t)$
except for its discrete dots.

Next, we think about random sequences uniformly distributed in $[0,2\pi)$.
At first, we study some properties of sequences of real numbers,
which are uniformly distributed modulo $2\pi$
(but need not be random).
We begin by giving the following set ${\cal S}_{N}$,
which consists of $(N+1)$ elements,
\begin{equation}
{\cal S}_{N}=\{x_{j}:0\leq x_{j}<2\pi,j\in\{0,1,...,N\}\}.
\label{definition-uniform-distribution-01}
\end{equation}
Rewriting the elements of ${\cal S}_{N}$ as an ordered list
$(x_{0}, x_{1}, ..., x_{N})$,
we can consider it to be a real sequence.
A necessary and sufficient condition for the real sequence $(x_{0}, x_{1}, ..., x_{N})$
(or the set ${\cal S}_{N}$) to be uniformly distributed
in the range of $[0,2\pi)$ under the limit of $N\to +\infty$
is given as follows \cite{Coppel2009,Weyl1916,Davenport1963,Kuipers1974,Walkden2011}:
\begin{equation}
\lim_{N\to+\infty}\frac{1}{N+1}
\sum_{j=0}^{N}
\exp(imx_{j})
=
0
\quad\quad
\mbox{$\forall m\in\{\pm 1,\pm 2,...\}$}.
\label{Weyl-criterion-without-proof-0}
\end{equation}
These relations are called the Weyl criterion.

Let us consider a concrete example of ${\cal S}_{N}$ as follows:
\begin{equation}
{\cal S}_{N}=\{n\Delta t \pmod{2\pi}:n\in\{0,1,2,...,N\}\},
\label{modulo-2pi-random-sequence-0}
\end{equation}
where $\Delta t$ is a positive constant.
Substitution of Eq.~(\ref{modulo-2pi-random-sequence-0})
into the Weyl criterion given by Eq.~(\ref{Weyl-criterion-without-proof-0}) yields
\begin{equation}
\lim_{N\to+\infty}\frac{1}{N+1}
\sum_{n=0}^{N}
e^{imn\Delta t}
=
\lim_{N\to+\infty}\frac{1}{N+1}
\frac{1-e^{im(N+1)\Delta t}}{1-e^{im\Delta t}}.
\label{modulo-2pi-random-sequence-1}
\end{equation}
If $m\Delta t \bmod{2\pi} \neq 0$ $\forall m\in\{\pm 1,\pm 2,...\}$,
the right-hand side of Eq.~(\ref{modulo-2pi-random-sequence-1}) is equal to zero.

Hence, at least, letting $\Delta t$ be written as
\begin{equation}
\Delta t=\sigma\frac{p}{q},
\label{modulo-2pi-random-sequence-2-dash}
\end{equation}
where arbitrary positive integers $p$ and $q$ are coprime,
and $\sigma$ is unity or an arbitrary positive
irrational but not transcendental number,
we consider $\lim_{N\to+\infty}{\cal S}_{N}$ to be uniformly distributed for $[0,2\pi)$.
To derive this result, we make use of facts that $\pi$ is a transcendental number
and a multiple of $2\pi$ is never equal to
\begin{equation}
n\Delta t=n\sigma\frac{p}{q},
\label{modulo-2pi-random-sequence-3-dash}
\end{equation}
where $n$ is an arbitrary positive integer.
In fact, not only $\Delta t$ given by Eq.~(\ref{modulo-2pi-random-sequence-2-dash})
but also a large variety of other time displacements lets $\lim_{N\to+\infty}{\cal S}_{N}$
given by Eq.~(\ref{modulo-2pi-random-sequence-0}) be uniformly distributed for $[0,2\pi)$.
However, throughout this paper, we concentrate
on $\Delta t$ given by Eq.~(\ref{modulo-2pi-random-sequence-2-dash}).

From the above considerations, we can conclude that the sequence
$\lim_{N\to \infty}{\cal S}_{N}$ such that Eq.~(\ref{modulo-2pi-random-sequence-0})
is uniformly distributed modulo $2\pi$,
where $\Delta t$ is given by Eq.~(\ref{modulo-2pi-random-sequence-2-dash}).
Moreover, we pay attention to the following fact.
If we assume
\begin{equation}
\pi<\Delta t<2\pi,
\label{condition- pseudorandom-Delta-t-0}
\end{equation}
we can consider $\lim_{N\to+\infty}{\cal S}_{N}$ to be a pseudorandom sequence
uniformly distributed for $[0,2\pi)$.
This is because a relation $2\Delta t > 2\pi$ generates an effect of the linear congruence method on
the sequence $n\Delta t$ modulo $2\pi$ for $n=0, 1, 2, ...$.

We can confirm this effect by giving a concrete counter-example for $0<\Delta t\ll \pi$.
For example, taking $\Delta t=0.05$, we obtain
\begin{equation}
125\Delta t=6.25,
\quad\quad
126\Delta t=6.3,
\quad\quad
125\Delta t< 2\pi < 126\Delta t,
\end{equation}
so that $(0,\Delta t,2\Delta t,...,125\Delta t)$ form a finite arithmetic progression.
Elements of the progression do not suffer from an effect of a constant modulus $2\pi$.
When we reach at $126\Delta t$, we observe an effect of the linear congruence method first.
In this case, evidently, we cannot think $\lim_{N\to+\infty}{\cal S}_{N}$ to be a random sequence.

Using the results obtained above,
we discuss an invariance of $S_{z}(t)[=L_{4}(t)]$ under a change of the scale $\Delta t$.
Here, we introduce a variable $b=e^{-\beta}$.
Because of $0<\beta<+\infty$, we obtain $0<b<1$.
Thus, we can rewrite $L_{4}(t)$ given by Eq.~(\ref{definition-L1-L3-L4}) as
\begin{equation}
L_{4}(t)=-\frac{1}{2}(1-b)+\frac{b}{2}(1-\frac{1}{b})^{2}f(t),
\label{definition-L4-dash}
\end{equation}
\begin{equation}
f(t)=\sum_{n=1}^{\infty}b^{n}\cos(2\sqrt{n}t).
\label{definition-f}
\end{equation}
In the following paragraphs, we show that the function $f(t)$ given by Eq.~(\ref{definition-f})
has a special scaling property.

First, we consider a set ${\cal S}$ of pseudorandom numbers (or a sequence)
distributed uniformly for $[0,2\pi)$
as follows:
\begin{equation}
{\cal S}=\{t_{n} \pmod{2\pi}:n\in\{0,1,2,...\}\}.
\label{set-S-scaling-01}
\end{equation}
Substitution of elements of ${\cal S}$ into $f(t)$ defined in Eq.~(\ref{definition-f})
yields a set of ${\cal F}$ as
\begin{equation}
{\cal F}=\{f(t_{0}), f(t_{1}), f(t_{2}),...\}.
\label{set-F-scaling-01}
\end{equation}
Here, let us consider an arbitrary real but not transcendental number $s(>1)$
and generate the following set:
\begin{equation}
{\cal S}'=\{st_{n} \pmod{2\pi}:n\in\{0,1,2,...\}\}.
\label{set-S-scaling-02}
\end{equation}
Moreover, we construct a new set ${\cal F}'$ from ${\cal S}'$ as follows:
\begin{equation}
{\cal F}'=\{f(st_{0}), f(st_{1}), f(st_{2}),...\}.
\label{set-F-scaling-02}
\end{equation}
Then, we cannot distinguish ${\cal S}$ from ${\cal S}'$
or ${\cal F}$ from ${\cal F}'$ actually on condition that both sets ${\cal S}$ and ${\cal S}'$
are countably infinite and their cardinalities are equal to each other,
that is, $|{\cal S}|=|{\cal S}'|$.
We argue this statement in the following paragraphs.

First, we think about $\Delta t$ given by
\begin{equation}
\Delta t=\frac{p}{q},
\quad\quad
\frac{\pi}{2}< \Delta t<\pi,
\end{equation}
where arbitrary positive integers $p$ and $q$ are coprime.
For example, we can choose $\Delta t=7/4=1.75$.
From the rational number $\Delta t$,
we generate a set as follows:
\begin{equation}
{\cal S}_{M,\Delta t}=\{m\Delta t \pmod{\pi}:m\in\{0,1,2,...,M\}\}.
\end{equation}
As explained before,
if we take sufficiently large $M$,
we can consider ${\cal S}_{M,\Delta t}$ to be uniformly distributed for $[0,\pi)$.
In addition, because of the effect of the constant modulus $\pi$,
we can regard ${\cal S}_{M,\Delta t}$ as a sequence of the pseudorandom numbers.
Moreover, replacing $\sigma$ with $\sqrt{n}$ in Eq.~(\ref{modulo-2pi-random-sequence-2-dash})
that gives the definition of $\Delta t$,
we obtain new pseudorandom sequences uniformly distributed for $[0,\pi)$,
\begin{eqnarray}
{\cal S}_{M,\sqrt{n}\Delta t}&=&\{m\sqrt{n}\Delta t \pmod{\pi}:m\in\{0,1,2,...,M\}\} \nonumber \\
&&\quad\quad
\mbox{for $n=1,2,3,...$}.
\label{random-sequence-sqrt-n}
\end{eqnarray}

Here, using the function $f(t)$ given by Eq.~(\ref{definition-f}),
let us construct a set,
\begin{equation}
{\cal F}_{M,\Delta t}=\{f(m\Delta t \pmod{\pi}):m\in\{0,1,2,..,M\}\}.
\end{equation}
In Eq.~(\ref{definition-f}), the $n$th term of the series $f(t)$ is given by $b^{n}\cos(2\sqrt{n}t)$.
Thus, a set of arguments $\sqrt{n}t$ of this cosine function
is equivalent to ${\cal S}_{M,\sqrt{n}\Delta t}$ defined in Eq.~(\ref{random-sequence-sqrt-n}).

Furthermore, taking an arbitrary real but not transcendental number $s(>1)$,
we think about a change of the scale $\Delta t\to s\Delta t$.
This change of the scale yields the following sets:
\begin{eqnarray}
{\cal S}_{M,s\sqrt{n}\Delta t}&=&\{sm\sqrt{n}\Delta t \pmod{\pi}:m\in\{0,1,2,..,M\}\} \nonumber \\
&&\quad\quad
\mbox{for $n=1,2,3,...$},
\end{eqnarray}
\begin{equation}
{\cal F}_{M,s\Delta t}=\{f(sm\Delta t \pmod{\pi}):m\in\{0,1,2,..,M\}\}.
\end{equation}
In the limit of $M\to +\infty$,
both ${\cal S}_{M,\sqrt{n}\Delta t}$ and ${\cal S}_{M,s\sqrt{n}\Delta t}$
become pseudorandom sequences uniformly distributed $[0,\pi)$,
because $s\Delta t$ satisfies the definition of $\Delta t$
given by Eq.~(\ref{modulo-2pi-random-sequence-2-dash}).
This implies that we cannot distinguish
between ${\cal S}_{M,\sqrt{n}\Delta t}$ and ${\cal S}_{M,s\sqrt{n}\Delta t}$ actually
in the limit of $M\to +\infty$.
Thus, we cannot distinguish between ${\cal F}_{M,\Delta t}$ and ${\cal F}_{M,s\Delta t}$
actually under the limit of $M\to +\infty$, too.
Let us describe these results as
${\cal S}_{M,\sqrt{n}\Delta t}\simeq {\cal S}_{M,s\sqrt{n}\Delta t}$
and
${\cal F}_{M,\Delta t}\simeq {\cal F}_{M,s\Delta t}$.
From these discussions,
we can conclude as follows:
taking the change of the scale $\Delta t\to s\Delta t$ for $s>1$,
we observe the scale invariance, ${\cal S}\simeq {\cal S}'$ and ${\cal F}\simeq {\cal F}'$,
for Eqs.~(\ref{set-S-scaling-01}), (\ref{set-F-scaling-01}), (\ref{set-S-scaling-02})
and (\ref{set-F-scaling-02}).

We can find similar scale invariance for $S_{x}(t)[=L_{1}(t)]$.
Substituting $b=e^{-\beta}$ into Eq.~(\ref{definition-L1-L3-L4}),
we can rewrite $L_{1}(t)$ as
\begin{equation}
L_{1}(t)=(1-b)g(t),
\label{definition-L1-dash}
\end{equation}
\begin{equation}
g(t)=\sum_{n=0}^{\infty}b^{n}\cos(\sqrt{n+1}t)\cos(\sqrt{n}t).
\label{definition-g}
\end{equation}
Substitution of elements of ${\cal S}$ given by Eq.~(\ref{set-S-scaling-01})
into $g(t)$ yields a set of
\\
${\cal G}=\{g(t_{0}),g(t_{1}),g(t_{2}),...\}$.
Then, the set ${\cal G}$ also acquires scaling properties,
which are similar to ${\cal F}\simeq{\cal F}'$
for Eqs.~(\ref{set-F-scaling-01}) and (\ref{set-F-scaling-02}).

To provide pseudorandom sequences to functions $L_{4}(t)$ and $f(t)$
defined in Eqs.~(\ref{definition-L4-dash}) and (\ref{definition-f}),
we choose $(\pi/2)<\Delta t<\pi$.
However, if we think about $L_{1}(t)$ and $g(t)$ defined
in Eqs.~(\ref{definition-L1-dash}) and (\ref{definition-g}),
we have to choose $\pi<\Delta t<2\pi$ for generating pseudorandom sequences safely.
The reason why we have to be careful in adjusting $\Delta t$ is the difference of a numerical factor two
between arguments of the $n$th terms of series $f(t)$ and $g(t)$,
that is,
$\cos(2\sqrt{n}t)$ and $\cos(\sqrt{n+1}t)\cos(\sqrt{n}t)$.
Hence, we set $\Delta t=7/2=3.5$ for Figs.~\ref{Figure02}, \ref{Figure03}, \ref{Figure04},
\ref{Figure05} and \ref{Figure06}.

In Appendix~\ref{section-appendix-A},
we examine physical transient spectra of the atom in the cavity.
Moreover, to understand the physical meanings and the scale invariance
of the discrete plots of the trajectories of the Bloch vector,
we investigate difference between continuous and discrete Fourier transforms for the atomic fluorescence.
To confirm the scale invariance of the discrete plots,
we examine histograms of the samples of the atomic fluorescence at constant time intervals.

\section{\label{section-L4-zero-points-numerical-approach}
The graph of $S_{x}(t)$ versus the inverse of the temperature $\beta$ for
the time $t$ such that $|S_{z}(t)|\ll 1$:
the numerical experiments}
Figures~\ref{Figure02}, \ref{Figure03} and \ref{Figure04} suggest to us
that distribution of points with $|S_{z}|\ll 1$
depends on $\beta$ strongly.
More precisely, on the one hand,
for $\beta\gg 1$,
points of $|S_{z}|\ll 1$ are localized around $|S_{x}|\simeq 1$.
On the other hand, for $\beta\ll 1$,
points of $|S_{z}|\ll 1$ are spread over a range of $S_{x}\in [-1,1]$.
Figure~\ref{Figure07} illustrates this characteristic feature of the function
$L_{4}(t)[=S_{z}(t)]$ given by Eqs.~(\ref{definition-L4-dash}) and (\ref{definition-f}) well.
How to draw Fig.~\ref{Figure07} is as follows.
First, taking $0<\epsilon \ll 1$,
we look for $t_{n}(=n\Delta t)$,
each of which satisfies $|L_{4}(t_{n})|\leq\epsilon$.
Second, we plot points $(\beta,S_{x}(t_{n}))$
for these times $t_{n}$.

First of all, we show that the following relation holds,
\begin{equation}
L_{4}(t)=0
\Leftrightarrow
\cos(2\sqrt{n}t)=1
\quad\forall n\in\{1,2,3,...\}.
\label{equations-for-zeropoints-L4-01}
\end{equation}
We can derive Eq.~(\ref{equations-for-zeropoints-L4-01}) as follows.
On the one hand,
if we assume the right-hand statement of Eq.~(\ref{equations-for-zeropoints-L4-01}) holds,
we can rewrite Eq.~(\ref{definition-f}) as
\begin{equation}
f(t)=\sum_{n=1}^{\infty}b^{n}=\frac{b}{1-b}.
\label{function-f-1}
\end{equation}
Substitution of Eq.~(\ref{function-f-1}) into Eq.~(\ref{definition-L4-dash}) yields
$L_{4}(t)=0$.
On the other hand, taking care of $0<b<1$ and $-1\leq\cos(2\sqrt{n}t)\leq 1$ $\forall n\in\{1,2,3,...\}$
in Eq.~(\ref{definition-f}),
we can conclude $f(t)=b/(1-b)$ if and only if the right-hand statement
in Eq.~(\ref{equations-for-zeropoints-L4-01}) holds.
Thus, from these discussions,
we arrive at $L_{4}(t)=0$ if and only if $\cos(2\sqrt{n}t)=1$ $\forall n\in\{1,2,3,...\}$.

Therefore, we can consider the right-hand statement of Eq.~(\ref{equations-for-zeropoints-L4-01})
to be a necessary and sufficient condition for $L_{4}(t)=0$.
Then, we can rewrite this necessary and sufficient condition as
\begin{eqnarray}
2t&=&2n_{1}\pi, \nonumber \\
2\sqrt{2}t&=&2n_{2}\pi, \nonumber \\
2\sqrt{3}t&=&2n_{3}\pi, \nonumber \\
&&...,
\end{eqnarray}
where $n_{1}, n_{2}, n_{3}, ...$ are integers.
Thus, if we assume $n_{1}\neq 0$,
we obtain
\begin{equation}
\sqrt{2}=\frac{n_{2}}{n_{1}},
\quad\quad
\sqrt{3}=\frac{n_{3}}{n_{1}},
\quad\quad
... .
\label{square-root-conditions}
\end{equation}
Equation~(\ref{square-root-conditions}) insists that $\sqrt{2}, \sqrt{3}, ...$
are rational numbers,
so that this result causes a contradiction.
Hence, we achieve a conclusion that $L_{4}(t)=0$ if and only if $t=0$.

However, as mentioned in Sec.~\ref{section-quasiperiodicity},
it is possible that $|L_{4}(t)|\ll 1$ holds for some $t(>0)$.
Thus, let us think about a problem whether or not we can find $t$,
which satisfies $|L_{4}(t)|\leq\epsilon$ for $0<\exists\epsilon\ll 1$.
Before we try a treatment of an algebraic analysis,
we examine this problem with numerical calculations.
First, we choose an arbitrary rational number as a time displacement $\pi<\Delta t<2\pi$.
Second, taking a sequence of the time variable,
$t_{n}=n\Delta t$
for $n\in\{0,1,2,...,N\}$,
we construct a set,
\begin{equation}
\{L_{4}(t_{n}):n\in\{0,1,2,...,N\}\}.
\label{set-L4-tn-0}
\end{equation}
Third, choosing $0<\epsilon\ll 1$,
we gather $t_{n}$,
each of which satisfies $|L_{4}(t_{n})|\leq\epsilon$.
Because of Eqs.~(\ref{definition-L1-L3-L4}) and (\ref{Bloch-xz-plane}),
we can consider $L_{1}(=S_{x})$ to be a function
of $t_{n}$ and $\beta$.
Fourth, we plot $(\beta,L_{1}(t_{n}))$ for these times $t_{n}$ and obtain a graph of Fig.~\ref{Figure07}.
The graph of Fig.~\ref{Figure07} represents a dependence of $S_{x}$ on $\beta$
under the condition $|S_{z}|\leq\epsilon$.

When we produce the graph of Fig.~\ref{Figure07} with numerical calculations,
we have to pay attention to the following facts.
In general, if we fix $\beta$ and $N$ to certain values,
the number of $t_{n}$ such that $|L_{4}(t_{n})|\leq\epsilon$ decreases rapidly
as $\epsilon(>0)$ becomes smaller.
Thus, with fixing $\beta$,
we have to let $N$ be larger as $\epsilon(>0)$ becomes smaller and approaches to zero.
According to Eq.~(\ref{set-L4-tn-0}),
we let $L_{4}(t_{0})$, $L_{4}(t_{1})$, $L_{4}(t_{2})$, ..., $L_{4}(t_{N})$
form a set of samples, whose number of elements is given by $(N+1)$ for a certain fixed $\beta$.
Among the samples given by Eq.~(\ref{set-L4-tn-0}),
the number of $t_{n}$ with $|L_{4}(t_{n})|\leq\epsilon$ decreases rapidly as $\beta$ becomes smaller.
Thus, to let the number of $t_{n}$ with $|L_{4}(t_{n})|\leq\epsilon$ remain constant,
we have to cause $N$ to be larger as $\beta$ becomes smaller.
Because of these circumstances,
for actual numerical calculations,
we introduce the following relation,
\begin{equation}
N=N_{0}e^{c_{1}/\beta},
\label{assumption-N-01}
\end{equation}
where $N_{0}=619.3$ and $c_{1}=13.37$.
We apply Eq.~(\ref{assumption-N-01}) to calculations with $\epsilon=7.5\times 10^{-4}$
for $1.0\leq\beta\leq 5.0$.
According to Eq.~(\ref{assumption-N-01}), we obtain
$N=8978$ for $\beta=5.0$ and $N=396\mbox{ }659\mbox{ }904$ for $\beta=1.0$.

However, if we apply Eq.~(\ref{assumption-N-01}) to a case of $0.5\leq\beta<1.0$,
$N$ becomes too large and we cannot carry out numerical calculations actually.
Thus, for $0.5\leq\beta<1.0$, we put $N=4\times 10^{8}$ as a constant
and introduce an alternative relation,
\begin{equation}
\epsilon=\epsilon_{0}e^{c_{2}/\beta},
\label{assumption-epsilon-01}
\end{equation}
where $\epsilon_{0}=1.875\times 10^{-4}$ and $c_{2}=\ln 4$.
According to Eq.~(\ref{assumption-epsilon-01}),
we change $\epsilon$ as a function of $\beta$,
so that we obtain $\epsilon=7.5\times 10^{-4}$ for $\beta=1.0$
and $\epsilon=3.0\times 10^{-3}$ for $\beta=0.5$.

Following the above prescriptions,
we plot points of $(\beta,S_{x}(t_{n}))$ that satisfy $|S_{z}(t_{n})|\leq\epsilon$
for $0.5\leq\beta\leq 5.0$ in Fig.~\ref{Figure07}.
For all points plotted in Fig.~\ref{Figure07}, we set $\Delta t=3.5$.
In Fig.~\ref{Figure07}, the horizontal and vertical axes are scaled
logarithmically and linearly, respectively.
Turning our eyes on Fig.~\ref{Figure07},
we observe that distinctive curves come into existence at $S_{x}=\pm 1$
in the limit of $\beta\to +\infty$
and their branches grow out from their trunk and spread as $\beta$ becomes smaller.
In Sec.~\ref{section-perturbative-approach},
we analyse properties of these curves using perturbative techniques.

\section{\label{section-perturbative-approach}
The graph of $S_{x}(t)$ versus the inverse of the temperature $\beta$
for the time $t$ such that$|S_{z}(t)|\ll 1$:
the perturbative evaluation}
In this section, we examine values of the time $t$ that satisfy $|L_{4}(t)|\ll 1$
in perturbation theory.
In Sec.~\ref{section-quasiperiodicity},
we suggest that Fig.~\ref{Figure07} is an appearance of incommensurate angular frequencies
under the quasiperiodic dynamics.
Then, we explain how to obtain approximations of their ratio with continued fractions.

In Sec.~\ref{section-L4-zero-points-numerical-approach},
we obtain a result that
$L_{4}(t)=0$ holds if and only if $t=0$.
Let us regard $L_{4}(t)$ as a power series in $b$
from Eqs.~(\ref{definition-L4-dash}) and (\ref{definition-f}).
Then, we take $0<\epsilon\ll 1$ and investigate values of the time
variable $t$ which satisfy $|L_{4}(t)|\leq\epsilon$.
Thus, we begin the perturbation method with the following relation:
\begin{equation}
\frac{b}{1-b}-\frac{2\epsilon b}{(1-b)^{2}}
\leq \sum_{n=1}^{\infty}b^{n}\cos(2\sqrt{n}t) < \frac{b}{1-b}.
\label{arbitrary-epsilon-inequality-0}
\end{equation}
Here, taking the low temperature limit $\beta\gg 1$,
that is, $0<b\ll 1$,
we investigate Eq.~(\ref{arbitrary-epsilon-inequality-0}) in terms of the parameter $b$,
such that the zero, first etc., powers of $b$
correspond to the zero, first, etc., orders of the perturbation calculation.

\subsection{\label{subsection-b-first-and-second-order-perturbation}
The first and second-order perturbations}
At first, we discuss the first-order perturbation.
In first order, we can rewrite Eq.~(\ref{arbitrary-epsilon-inequality-0}) as
\begin{equation}
b-2\epsilon b\leq b\cos(2t)<b.
\label{first-order-perturbation-0}
\end{equation}
Taking the limit of $\epsilon\to +0$ for Eq.~(\ref{first-order-perturbation-0}),
we obtain $t=0,\pi,2\pi,3\pi,...$.
These results are not important for us indeed,
because they do not give us any physical meanings.

Next, we discuss the second-order perturbation.
In second order,
we can rewrite Eq.~(\ref{arbitrary-epsilon-inequality-0}) as
\begin{equation}
0<b[1-\cos(2t)]+b^{2}[1-\cos(2\sqrt{2}t)]\leq 2\epsilon b(1+2b).
\label{second-order-perturbation-0}
\end{equation}
It is very difficult for us to find all values of $t$,
each of which satisfies Eq.~(\ref{second-order-perturbation-0}).
Thus, giving up our attempts to find all $t$ with Eq.~(\ref{second-order-perturbation-0}),
we concentrate on specifying a subset of $t$ such that Eq.~(\ref{second-order-perturbation-0}).

Because of
$\cos(2t)\leq 1$
and
$\cos(2\sqrt{2}t)\leq 1$,
both the first and second terms of Eq.~(\ref{second-order-perturbation-0}) have to be
larger than or equal to zero.
Hence, let us consider a special case where the following two relations hold,
\begin{eqnarray}
b[1-\cos(2t)]&=&2\epsilon b, \label{case1-a} \\
b^{2}[1-\cos(2\sqrt{2}t)]&=&4\epsilon b^{2}. \label{case1-b}
\end{eqnarray}
We have to emphasize that some $t$ with Eq.~(\ref{second-order-perturbation-0})
may not satisfy Eqs.~(\ref{case1-a}) and (\ref{case1-b}).
However, we neglect this possibility throughout this section,
and we give all attention on Eqs.~(\ref{case1-a}) and (\ref{case1-b}).

Because Eq.~(\ref{case1-a}) is essentially equivalent to Eq.~(\ref{first-order-perturbation-0}),
we obtain $t=q\pi$ for $q=0, 1, 2, ...$.
Here, we remember that we obtain $q=t=0$ as a trivial root of $L_{4}(t)=0$
in Sec.~\ref{section-L4-zero-points-numerical-approach}.
Thus, we think about $q=1, 2, 3, ...$ only.
Then, from Eq.~(\ref{case1-b}), we obtain
\begin{equation}
\cos(2\sqrt{2}q\pi)=1-4\epsilon.
\label{second-order-perturbation-2}
\end{equation}
This implies the following fact.
There are a countably infinite number of inequalities,
\begin{eqnarray}
|2\sqrt{2}q\pi-2p\pi|&<&\delta(4\epsilon)\ll 1, \label{case1-condition-1} \\
|2\sqrt{2}q\pi-4p\pi|&<&\delta(4\epsilon)\ll 1, \label{case1-condition-2} \\
|2\sqrt{2}q\pi-6p\pi|&<&\delta(4\epsilon)\ll 1, \label{case1-condition-3} \\
|2\sqrt{2}q\pi-8p\pi|&<&\delta(4\epsilon)\ll 1, \label{case1-condition-4} \\
&...& \nonumber,
\end{eqnarray}
where an explicit form of $\delta(4\epsilon)$ is given in a next paragraph.
Among the above inequalities,
only one inequality holds for a certain positive integer $p$,
where $p$ and $q$ are coprime.

An explicit form of a function $\delta(4\epsilon)$ appearing
in Eqs.~(\ref{case1-condition-1}), (\ref{case1-condition-2}),
(\ref{case1-condition-3}) and (\ref{case1-condition-4}) is given by
\begin{equation}
\delta(4\epsilon)=|\arccos(1-4\epsilon)|.
\label{function-delta-epsilon-1}
\end{equation}
Equation~(\ref{function-delta-epsilon-1}) tells us that $\delta(4\epsilon)$
is equal to a very small positive number.
From Eq.~(\ref{case1-condition-1}),
we can derive the following inequality:
\begin{equation}
|\sqrt{2}\frac{q}{p}-1|<\frac{\delta(4\epsilon)}{2\pi p}
\ll 1.
\label{case1-condition-1-a}
\end{equation}
Thus, we can conclude
$p/q\simeq\sqrt{2}$,
so that we arrive at the fact that
$p/q$ is a rational approximate number of $\sqrt{2}$ finally.
In other words, one of the time variables with Eqs.~(\ref{case1-a}) and (\ref{case1-b})
is given by $t=q\pi$,
where $p/q$ is a rational approximate number of $\sqrt{2}$.

Then, the following question arises.
How small is the upper bound of $|\sqrt{2}-(p/q)|$?
How do we estimate the numerical precision of $p/q$
as an approximate number of $\sqrt{2}$?
We discuss the question of accuracy later.

From Eq.~(\ref{case1-condition-2}), we obtain
\begin{equation}
|\frac{1}{\sqrt{2}}\frac{q}{p}-1|<\frac{\delta(4\epsilon)}{4\pi p}
\ll 1.
\label{case1-condition-2-a}
\end{equation}
This inequality yields a conclusion of
$p/q\simeq 1/\sqrt{2}$, so that
we arrive at the fact that $p/q$ is a rational approximate number of $1/\sqrt{2}$.
This implies that one of the time variables with Eqs.~(\ref{case1-a}) and (\ref{case1-b})
is given by $t=q\pi$, where $p/q$ is a rational approximate number of $1/\sqrt{2}$.

Applying a similar discussion to Eq.~(\ref{case1-condition-3}), we obtain
$p/q\simeq\sqrt{2}/3$ and $t=q\pi$.
Moreover, applying a similar discussion to Eq.~(\ref{case1-condition-4}),
we obtain
$p/q\simeq 1/(2\sqrt{2})$ and $t=q\pi$.
Putting these results together,
we can derive a general formula for $k=1,2,3,...$,
\begin{equation}
\frac{p}{q}\simeq\frac{\sqrt{2}}{k},
\quad
t=q\pi
\quad
\mbox{for $k=1,2,3,...$}.
\label{case1-condition-k-b}
\end{equation}

As a result of the above discussions,
we obtain the following conclusion.
In the second-order perturbation theory,
the time variable $t=q\pi$ satisfies $|L_{4}(t)|\ll 1$,
where $p/q$ is a rational approximate number of $\sqrt{2}/k$ for $k=1,2,3,...$,
and $p$ and $q$ are coprime.
Going into details,
a set of the time variables $t$ with $0\leq |L_{4}(t)|\leq\epsilon(\ll 1)$
includes the following elements $t=q\pi$.
First, we consider rational approximations $p/q$ of $\sqrt{2}/k$ for $k=1,2,3,...$.
Second, using a very small positive value $\delta(4\epsilon)$,
we can give their accuracies as
\begin{equation}
|\frac{\sqrt{2}}{k}\frac{q}{p}-1|<\frac{\delta(4\epsilon)}{2k\pi p}\ll 1
\quad\quad
\mbox{for $k=1,2,3,...$}.
\label{approximation-rational-number-1}
\end{equation}

We can rewrite Eq.~(\ref{approximation-rational-number-1}) as follows:
\begin{equation}
|\frac{\sqrt{2}}{k}-\frac{p}{q}|<\frac{\delta(4\epsilon)}{2k\pi q}\ll 1
\quad\quad
\mbox{for $k=1,2,3,...$}.
\label{approximation-rational-number-2}
\end{equation}
In the above equation, $p/q$ (where $q>0$) represents a rational approximate number of $\sqrt{2}/k$.
If we put $t=q\pi$ with Eq.~(\ref{approximation-rational-number-2}),
$0\leq |L_{4}(t)|\leq\epsilon$ holds.
Therefore, Eq.~(\ref{approximation-rational-number-2}) gives an upper bound of the error
resulting from this approximation.
Let us estimate the right-hand side of Eq.~(\ref{approximation-rational-number-2})
at $O(1/q)$.
Then, Eq.~(\ref{approximation-rational-number-2}) implies
that the precision of the rational number approximation $p/q$ to $\sqrt{2}/k$
has to be less than $O(1/q)$.

From the above considerations, we obtain the following result:
\begin{equation}
|\frac{\sqrt{2}}{k}-\frac{p}{q}|=\frac{1}{cq^{1+\nu}}
\quad\quad
\mbox{for $k=1,2,3,...$},
\label{sqrt-2-slash-k-approximated-rational-number-2}
\end{equation}
where $p$ and $q(>0)$ are coprime and $\nu>1$.
In addition, we suppose that $c$ is a constant and close to unity.
Furthermore, we want to let $\nu$ be as large as possible we can.
This is because a rational approximate number $p/q$ becomes closer to $\sqrt{2}/k$
as $\nu(>1)$ gets larger and larger.

Here, we make use of the following theorem,
which is related to the Diophantine approximation
\cite{Coppel2009,Niven1960}.
For an arbitrary irrational number $\alpha$,
there exist infinite sequences $p_{n}$ and $q_{n}(>0)$ for $n \geq 0$
such that $p_{n}$ and $q_{n}$ are coprime and
\begin{equation}
|q_{n}\alpha -p_{n}|<\frac{1}{q_{n}}.
\end{equation}
In other words, there exist infinitely many rational numbers $p/q$ such that
\begin{equation}
|\alpha-\frac{p}{q}|<\frac{1}{q^{2}},
\end{equation}
where $p$ and $q(>0)$ are coprime.
Moreover, the following fact is very useful for us in the remainder of this section.
For an arbitrary irrational but not transcendental number $\alpha$,
if we suppose that there exist infinitely many rational numbers $p/q$ such that
\begin{equation}
|\alpha-\frac{p}{q}|<\frac{1}{q^{\mu}},
\label{Diopantine-approximation-0}
\end{equation}
the upper bound of $\mu$ is equal to two.
This fact is called the theorem of Roth,
which is a fundamental result in the Diophantine approximation
\cite{Coppel2009}.
Because to prove this theorem is beyond our purpose of this paper,
we do not touch it any more.
Hence, from now on, we consider only a rational approximate number
$p/q$ of $\sqrt{2}/k$ for $k=1,2,3,...$
such that
\begin{equation}
|\frac{\sqrt{2}}{k}-\frac{p}{q}|<\frac{1}{q^{2}}
\quad\quad
\mbox{for $k=1,2,3,...$},
\label{sqrt-2-approximation-2}
\end{equation}
and we choose $t=q\pi$ for Eqs.~(\ref{case1-a}) and (\ref{case1-b}).

In the following paragraphs,
we confirm the above discussions with numerical calculations actually.
By computing first some terms in a continued fraction representation of an irrational number,
we can efficiently obtain its rational approximate number with high precision.
Let us describe the continued fraction of an arbitrary irrational number $\alpha$ as
\begin{eqnarray}
\alpha
&=&
a_{0}
+\frac{1}{\displaystyle a_{1}
+\frac{1}{\displaystyle a_{2}+ \frac{1}{\ddots}}} \nonumber \\
&=&
[a_{0};a_{1},a_{2},...],
\label{continued-fraction-definition-0}
\end{eqnarray}
where $a_{0}$ is an integer and $a_{1}, a_{2}, ...$
are positive integers.

The continued fractions of irrational numbers have the following properties,
whose proofs are given in Refs.~\cite{Coppel2009,Niven1960}.
\begin{enumerate}
\item
If a real number $\alpha$ is irrational,
its continued fraction expression
$[a_{0}; a_{1}, a_{2}, ...]$
is infinite.
\item
If and only if $\alpha$ is an irrational solution of a quadratic equation with rational coefficients,
its continued fraction expression is periodic.
\item
For an arbitrary irrational number
$\alpha=[a_{0};a_{1},a_{2},a_{3},...]$,
let us consider infinite sequences of integers
$(p_{0}, p_{1}, p_{2}, ...)$ and $(q_{0}, q_{1}, q_{2}, ...)$
such that
\begin{equation}
[a_{0};a_{1},a_{2},...,a_{n}]=\frac{p_{n}}{q_{n}}.
\label{continued-fraction-definition-finite-1}
\end{equation}
Obviously, from Eq.~(\ref{continued-fraction-definition-0}),
we can write down $p_{n}$ and $q_{n}$ as follows:
\begin{equation}
p_{-1}=1,
\quad
p_{0}=a_{0},
\quad
p_{n}=a_{n}p_{n-1}+p_{n-2},
\end{equation}
\begin{equation}
q_{-1}=0,
\quad
q_{0}=1,
\quad
q_{n}=a_{n}q_{n-1}+q_{n-2}.
\end{equation}
Then, the following relation holds,
\begin{equation}
\frac{p_{0}}{q_{0}}
<
\frac{p_{2}}{q_{2}}
<
\frac{p_{4}}{q_{4}}
<...<
\frac{p_{5}}{q_{5}}
<
\frac{p_{3}}{q_{3}}
<
\frac{p_{1}}{q_{1}}.
\end{equation}
\item
For a continued fraction expression of an arbitrary irrational number
\\
$\alpha=[a_{0};a_{1}, a_{2}, a_{3}, ...]$,
let us consider a quantity obtained by including its first $(n+1)$ terms
as shown in Eq.~(\ref{continued-fraction-definition-finite-1}).
Then the following inequalities hold,
\begin{equation}
|\alpha-\frac{p_{n}}{q_{n}}|\leq\frac{1}{a_{n+1}q_{n}^{2}}
\quad\quad
\mbox{for $n=0,1,2,...$}.
\end{equation}
\end{enumerate}

Let us obtain a rational approximate number of $\sqrt{2}/k$ for $k=1, 2, 3, ...$
by truncating its corresponding continued fraction.
We describe the obtained rational approximate number as $p/q$.
Then, because of the fourth item written above,
it satisfies Eq.~(\ref{sqrt-2-approximation-2}).
[Here, we pay attention to the following fact.
In general, we cannot find all rational approximate numbers of $\sqrt{2}/k$
with Eq.~(\ref{sqrt-2-approximation-2})
by using the continued fraction.]

The continued fraction of a quadratic irrational number $\sqrt{2}$
is given by
\begin{equation}
\sqrt{2}
=
[1;2,2,2,...]
=
[1;\dot{2}].
\label{continued-fraction-square-root-two-a}
\end{equation}
In Eq.~(\ref{continued-fraction-square-root-two-a}),
$a_{0}=1$ and $a_{1}=a_{2}=...=2$ hold,
so that the figure ``$2$" appears in succession with period unity.
From now on, we indicate the repeating block by dots as shown
in Eqs.~(\ref{continued-fraction-square-root-two-a}) and
(\ref{continued-fraction-square-roos}).
Moreover, we introduce the following notations:
\begin{eqnarray}
\chi_{\sqrt{2}}(0)
&=&
[1]=1, \nonumber \\
\chi_{\sqrt{2}}(1)
&=&
[1;2]=3/2, \nonumber \\
\chi_{\sqrt{2}}(2)
&=&
[1;2,2]=7/5, \nonumber \\
\chi_{\sqrt{2}}(n)
&=&
[1;\overbrace{2,...,2}^{n}].
\label{continued-fraction-sqrt-2-approximations-0}
\end{eqnarray}

Because $1/\sqrt{2}$, $\sqrt{2}/3$ and $1/(2\sqrt{2})$ are quadratic irrational numbers,
we can write down them in the following continued fraction expressions, as well:
\begin{eqnarray}
1/\sqrt{2}
&=&
[0;1,2,2,2,...]
=
[0;1,\dot{2}], \nonumber \\
\sqrt{2}/3
&=&
[0;2,8,4,8,4,8,4,...]
=
[0;2,\dot{8},\dot{4}], \nonumber \\
1/(2\sqrt{2})
&=&
[0;2,1,4,1,4,1,4,...]
=
[0;2,\dot{1},\dot{4}].
\label{continued-fraction-square-roos}
\end{eqnarray}
Moreover, we obtain
\begin{eqnarray}
&&
\chi_{1/\sqrt{2}}(0)=[0]=0,
\quad
\chi_{1/\sqrt{2}}(1)=[0;1]=1,
\quad
\chi_{1/\sqrt{2}}(2)=[0;1,2]=2/3, \nonumber \\
&&
\chi_{1/\sqrt{2}}(n)=[0;\overbrace{1,2,...,2}^{n}], \nonumber \\
&&
\chi_{\sqrt{2}/3}(0)=[0]=0,
\quad
\chi_{\sqrt{2}/3}(1)=[0;2]=1/2, \nonumber \\
&&
\chi_{\sqrt{2}/3}(2)=[0;2,8]=8/17, \nonumber \\
&&
\chi_{\sqrt{2}/3}(n)=[0;\overbrace{2,8,4,...,a_{n}}^{n}], \nonumber \\
&&
\chi_{1/(2\sqrt{2})}(0)=[0]=0,
\quad
\chi_{1/(2\sqrt{2})}(1)=[0;2]=1/2, \nonumber \\
&&
\chi_{1/(2\sqrt{2})}(2)=[0;2,1]=1/3, \nonumber \\
&&
\chi_{1/(2\sqrt{2})}(n)=[0;\overbrace{2,1,4,...,a_{n}}^{n}].
\end{eqnarray}

Let us consider Eq.~(\ref{sqrt-2-approximation-2}) for $k=1, 2, 3, 4$.
(We neglect the cases where $k=5, 6, 7, ...$.)
Then, we choose rational approximate numbers
of $\sqrt{2}$, $1/\sqrt{2}$, $\sqrt{2}/3$ and $1/(2\sqrt{2})$ as follows.
At first, thinking about rational approximate numbers of $\sqrt{2}$,
we dismiss $\chi_{\sqrt{2}}(0)$, ..., $\chi_{\sqrt{2}}(11)$
in order to obtain good accuracy.
Thus, we select the following $28$ rational numbers for the approximation of $\sqrt{2}$,
\begin{eqnarray}
&&
\chi_{\sqrt{2}}(12)=\frac{47\mbox{ }321}{33\mbox{ }461},
\quad
\chi_{\sqrt{2}}(13)=\frac{114\mbox{ }243}{80\mbox{ }782},
\quad
..., \nonumber \\
&&\quad
\chi_{\sqrt{2}}(39)
=
\frac{1\mbox{ }023\mbox{ }286\mbox{ }908\mbox{ }188\mbox{ }737}
{723\mbox{ }573\mbox{ }111\mbox{ }879\mbox{ }672}.
\label{approximation-rational-numbers-A}
\end{eqnarray}
We do not select $\chi_{\sqrt{2}}(40), \chi_{\sqrt{2}}(41), \chi_{\sqrt{2}}(42), ...$
because each number of digits in their denominators is larger than $15$
and they are not tractable in the numerical calculations actually.
In a similar way,
we obtain rational approximate numbers of $1/\sqrt{2}$, $\sqrt{2}/3$ and $1/(2\sqrt{2})$
as follows:
\begin{eqnarray}
&&\chi_{1/\sqrt{2}}(12)=\frac{13\mbox{ }860}{19\mbox{ }601},
\quad
\chi_{1/\sqrt{2}}(13)=\frac{33\mbox{ }461}{47\mbox{ }321},
\quad
..., \nonumber \\
&&\quad
\chi_{1/\sqrt{2}}(39)
=
\frac{299\mbox{ }713\mbox{ }796\mbox{ }309\mbox{ }065}{423\mbox{ }859\mbox{ }315\mbox{ }570\mbox{ }607},
\label{approximation-rational-numbers-B}
\end{eqnarray}
\begin{eqnarray}
&&
\chi_{\sqrt{2}/3}(12)=\frac{362\mbox{ }226\mbox{ }480}{768\mbox{ }398\mbox{ }401},
\quad
\chi_{\sqrt{2}/3}(13)=\frac{1\mbox{ }492\mbox{ }851\mbox{ }361}{3\mbox{ }166\mbox{ }815\mbox{ }962},
\quad
...,
\nonumber \\
&&\quad\quad
\chi_{\sqrt{2}/3}(19)
=
\frac{58\mbox{ }522\mbox{ }759\mbox{ }015\mbox{ }841}{124\mbox{ }145\mbox{ }519\mbox{ }261\mbox{ }542},
\label{approximation-rational-numbers-C}
\end{eqnarray}
\begin{eqnarray}
&&
\chi_{1/(2\sqrt{2})}(12)=\frac{6930}{19\mbox{ }601},
\quad
\chi_{1/(2\sqrt{2})}(13)=\frac{33\mbox{ }461}{94\mbox{ }642},
\quad
...,
\nonumber \\
&&\quad\quad
\chi_{1/(2\sqrt{2})}(39)
=
\frac{299\mbox{ }713\mbox{ }796\mbox{ }309\mbox{ }065}{847\mbox{ }718\mbox{ }631\mbox{ }141\mbox{ }214}.
\label{approximation-rational-numbers-D}
\end{eqnarray}

In fact, we can verify that the following relations hold,
\begin{eqnarray}
|\sqrt{2}-\chi_{\sqrt{2}}(12)|
&\simeq&
3.158\times 10^{-10}<1/(2\times 33\mbox{ }461^{2}), \nonumber \\
|\sqrt{2}-\chi_{\sqrt{2}}(13)|
&\simeq&
5.418\times 10^{-11}<1/(2\times 80\mbox{ }782^{2}),
\end{eqnarray}
and
\begin{eqnarray}
1-\cos(2\sqrt{2}\times 33\mbox{ }461\pi)
&\simeq&
2.204\times 10^{-9}, \nonumber \\
1-\cos(2\sqrt{2}\times 80\mbox{ }782\pi)
&\simeq&
3.781\times 10^{-10}.
\end{eqnarray}

\begin{figure}
\begin{center}
\includegraphics[scale=1.0]{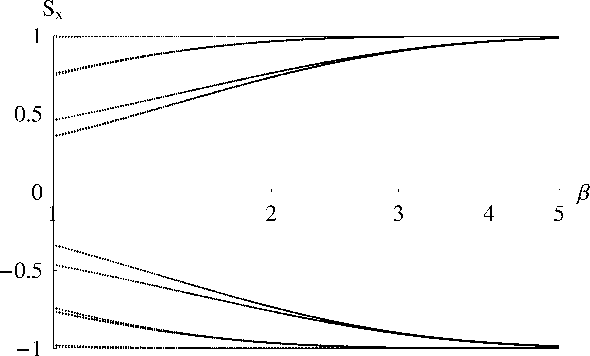}
\end{center}
\caption{A linear-log plot of
$(\beta,S_{x}(t))$ for $t=q\pi$ where $q\in{\cal M}$
given by Eq.~(\ref{integer-sets-zeropoints-0})
and $2.5\leq\beta\leq 5.0$,
and for $t=q\pi$ where $q\in\tilde{\cal M}$
given by Eq.~(\ref{set-tilde-M-third-order-0}) and $1.0\leq\beta\leq 2.5$.
All dots are drawn with changing $\beta$ at constant interval of $0.01$.
We give them a diameter being $4/1000$ of the width of the whole graph.}
\label{Figure11}
\end{figure}

From Eqs.~(\ref{approximation-rational-numbers-A}),
(\ref{approximation-rational-numbers-B}),
(\ref{approximation-rational-numbers-C}) and
(\ref{approximation-rational-numbers-D}),
we obtain $70$ different positive integers as denominators.
[Some integers appear as denominators twice or more
in Eqs.~(\ref{approximation-rational-numbers-A}),
(\ref{approximation-rational-numbers-B}),
(\ref{approximation-rational-numbers-C}) and
(\ref{approximation-rational-numbers-D}),
so that we have to avoid counting them over again.]
Putting these $70$ integers and zero together,
we construct a set ${\cal M}$,
whose cardinality is equal to $71$, as
\begin{eqnarray}
{\cal M}
&=&
\{
0,
19\mbox{ }601,
33\mbox{ }461,
47\mbox{ }321,
80\mbox{ }782,
94\mbox{ }642,
..., \nonumber \\
&&
\quad\quad
723\mbox{ }573\mbox{ }111\mbox{ }879\mbox{ }672, \nonumber \\
&&
\quad\quad
847\mbox{ }718\mbox{ }631\mbox{ }141\mbox{ }214
\}.
\label{integer-sets-zeropoints-0}
\end{eqnarray}
Then, we obtain $71$ distinct times,
$t=q\pi$
for $q\in {\cal M}$.
According to Eqs.~(\ref{definition-L1-L3-L4}) and (\ref{Bloch-xz-plane}),
we consider $S_{x}(=L_{1})$ to be a function of $t$ and $\beta$ and plot the points
of $(\beta,S_{x}(t))$ for $t=q\pi$ where $q\in {\cal M}$
and $2.5\leq\beta\leq 5.0$ in Fig.~\ref{Figure11}.
Figure~\ref{Figure11} reproduces the graph of Fig.~\ref{Figure07}
for $2.5\leq\beta\leq 5.0$ well.

\subsection{The third-order perturbation}
In this subsection, we discuss the third-order perturbation.
In third order, we can rewrite Eq.~(\ref{arbitrary-epsilon-inequality-0}) as
\begin{equation}
0<1-\tilde{f}(b,t)\leq\frac{2\epsilon b(1+2b+3b^{2})}{b+b^{2}+b^{3}},
\label{b-third-order-perturbation-0}
\end{equation}
\begin{equation}
\tilde{f}(b,t)=\frac{1}{b+b^{2}+b^{3}}
[b\cos(2t)+b^{2}\cos(2\sqrt{2}t)+b^{3}\cos(2\sqrt{3}t)].
\label{b-third-order-perturbation-1}
\end{equation}
After discussions given in Sec.~\ref{subsection-b-first-and-second-order-perturbation},
we give up our attempts to find all $t$
with Eqs.~(\ref{b-third-order-perturbation-0}) and (\ref{b-third-order-perturbation-1}).
Thus, we concentrate on specifying a subset of $t$, each of which satisfies
Eqs.~(\ref{b-third-order-perturbation-0}) and (\ref{b-third-order-perturbation-1}).
Obviously, the time $t$
with Eqs.~(\ref{b-third-order-perturbation-0}) and (\ref{b-third-order-perturbation-1})
has to satisfy Eq.~(\ref{second-order-perturbation-0}), as well.
Thus, it is possible that some of $t$
with Eqs.~(\ref{b-third-order-perturbation-0}) and (\ref{b-third-order-perturbation-1})
belong to the set of $t=q\pi$ for $q\in{\cal M}$ given by Eq.~(\ref{integer-sets-zeropoints-0}).
Hence, we try finding the time $t$
with Eqs.~(\ref{b-third-order-perturbation-0}) and (\ref{b-third-order-perturbation-1})
from $t=q\pi$ for $q\in{\cal M}$.

In fact, we select $q\in {\cal M}$ for $t=q\pi$
with Eqs.~(\ref{b-third-order-perturbation-0}) and (\ref{b-third-order-perturbation-1})
on the following condition.
Fixing the parameter for perturbation at $\epsilon\simeq 0.014\mbox{ }78$ and
$b=e^{-2.0}\simeq0.1353$,
that is, $\beta=2.0$,
and obtaining $2\epsilon b(1+2b+3b^{2})/(b+b^{2}+b^{3})\simeq0.002$,
we look for $q$ such that
\begin{equation}
1-\tilde{f}(e^{-2.0},q\pi)<0.002
\quad\quad
\mbox{for $q\in {\cal M}$}.
\label{b-third-order-perturbation-beta-2}
\end{equation}
As a result of numerical calculations,
we find $14$ integers for $q$ satisfying Eq.~(\ref{b-third-order-perturbation-beta-2}),
\begin{eqnarray}
\tilde{\cal M}&=&\{
0,
19\mbox{ }601,
470\mbox{ }832,
1\mbox{ }607\mbox{ }521, \nonumber \\
&&
15\mbox{ }994\mbox{ }428,
18\mbox{ }738\mbox{ }638,
768\mbox{ }398\mbox{ }401, \nonumber \\
&&
10\mbox{ }812\mbox{ }186\mbox{ }007,
21\mbox{ }624\mbox{ }372\mbox{ }014, \nonumber \\
&&
627\mbox{ }013\mbox{ }566\mbox{ }048,
8\mbox{ }822\mbox{ }750\mbox{ }406\mbox{ }821, \nonumber \\
&&
30\mbox{ }122\mbox{ }754\mbox{ }096\mbox{ }401,
299\mbox{ }713\mbox{ }796\mbox{ }309\mbox{ }065, \nonumber \\
&&
847\mbox{ }718\mbox{ }631\mbox{ }141\mbox{ }214
\}.
\label{set-tilde-M-third-order-0}
\end{eqnarray}
Thus, regarding $S_{x}(=L_{1})$ as a function of $t$ and $\beta$ because of
Eqs.~(\ref{definition-L1-L3-L4}) and (\ref{Bloch-xz-plane}),
we plot the points of
$(\beta,S_{x}(t))$ for $t=q\pi$ where $q\in\tilde{\cal M}$ of Eq.~(\ref{set-tilde-M-third-order-0}) and
$1.0\leq\beta\leq 2.5$ in Fig.~\ref{Figure11}.
We observe that Fig.~\ref{Figure11} reproduces the graph of Fig.~\ref{Figure07}
for $1.0\leq\beta\leq 2.5$ well.

Before closing this subsection,
we make two remarks concerning our treatments of the Diophantine approximation.
First, we indicate
a property of integers $q\in\tilde{\cal M}$ in Eq.~(\ref{set-tilde-M-third-order-0}).
Remembering Eq.~(\ref{sqrt-2-approximation-2}),
we expect $\forall q\in\tilde{\cal M}$ to satisfy
the following relation for certain positive integers $k$ and $q$:
\begin{equation}
|\frac{\sqrt{3}}{k}-\frac{p}{q}|<\frac{1}{q^{2}}.
\end{equation}
In fact, for example,
for $q=19\mbox{ }601\in\tilde{\cal M}$ and $q=470\mbox{ }832\in\tilde{\cal M}$,
we have the following two relations, respectively:
\begin{equation}
|\frac{\sqrt{3}}{2425}-\frac{14}{19\mbox{ }601}|<\frac{1}{19\mbox{ }601^{2}},
\end{equation}
\begin{equation}
|\frac{\sqrt{3}}{163\mbox{ }101}-\frac{5}{470\mbox{ }832}|<\frac{1}{470\mbox{ }832^{2}}.
\end{equation}

Second, we have to point out that values of the time co-ordinate $t=q\pi$
for $q\in{\cal M}$ and $q\in\tilde{\cal M}$ given by Eqs.~(\ref{integer-sets-zeropoints-0}) and
(\ref{set-tilde-M-third-order-0})
never fit in the distinct times $t_{n}=n\Delta t$, $\Delta t=p/q$ and $\pi<\Delta t<2\pi$
discussed in Sec.~\ref{section-scale-invariance}.
This implies that the graph of Fig.~\ref{Figure07} is plotted at times of rational numbers
and the graph of Fig.~\ref{Figure11} is plotted at times of irrational (and transcendental) numbers.
We cannot find an acceptable way of dealing this difference
between Figs.~\ref{Figure07} and \ref{Figure11}.

\section{\label{section-discussions}
Discussions}
We can obtain the JCM by applying the rotating-wave approximation to a single two-level atom
that interacts with a single mode of an optical cavity.
In this paper,
we investigate the quasiperiodicity,
which the atom and the cavity field show during the time evolution
according to the Jaynes-Cummings interaction,
where the atom and the cavity field are initially put into a certain pure state
and a mixed state in thermal equilibrium, respectively.

However, for actual experiments of the cavity quantum electrodynamics (QED) in the laboratory,
we may not observe quasiperiodicity because of lack of the rotating-wave approximation.
For example,
it is possible that we cannot reproduce robust experimental results of scale invariance,
which we discuss in this paper.

In Ref.~\cite{Milonni1983},
Milonni {\it et al.} describe the interaction between a collection of two-level atoms
and the single-mode classical electric field as the optical Bloch equations
using the semiclassical approximation.
They are given as follows:
\begin{eqnarray}
\dot{X}(t)
&=&
-\omega_{0}Y(t), \nonumber \\
\dot{Y}(t)
&=&
\omega_{0}X(t)
+
(2/\hbar)pE(t)Z(t) , \nonumber \\
\dot{Z}(t)
&=&
-(2/\hbar)pE(t)Y(t) , \nonumber \\
\ddot{E}(t)+\omega^{2}E(t)&=&-4\pi Np\ddot{X}(t),
\end{eqnarray}
where $(X(t),Y(t),Z(t))$ represents the Bloch vector,
$p$ is the transition dipole moment,
$E(t)$ represents the electric field,
$\omega_{0}$ and $\omega$ are angular frequencies of the atoms and the electric field respectively,
and $N$ is a density of the atoms.

Milonni {\it et al.} investigate the optical Bloch equations numerically
and obtain the following results.
In the rotating-wave approximation,
there is no predictions of chaos.
On the other hand,
if the initial conditions let the rotating-wave approximation fail,
we can occasionally obtain the chaotic behaviour.

In Ref.~\cite{Prants2002},
Prants {\it et al.} consider the recoil effect caused by the centre-of-mass motion of the atom
in the cavity QED.
They extend the Hamiltonian of the JCM as follows:
\begin{equation}
\hat{H}
=
\frac{\hat{p}^{2}}{2m}
+
\frac{\hbar}{2}\omega_{a}\hat{\sigma}_{z}
+
\hbar\omega_{f}\hat{a}^{\dagger}\hat{a}
-
\frac{\hbar\Omega_{0}}{2}(\hat{\sigma}_{+}\hat{a}+\hat{\sigma}_{-}\hat{a}^{\dagger})
\cos k_{f}\hat{x},
\label{extended-Hamiltonian-JCM-recoil}
\end{equation}
where $\hat{x}$ and $\hat{p}$ are the atomic position and momentum operators, respectively.
In the above Hamiltonian,
we consider the atom to be in a single-mode high-finesse standing-wave cavity,
so that the dynamics of the system is sensitive to the centre-of-mass motion of the atom.

Prants {\it et al.} apply the semiclassical approximation to this extended JCM.
They assume that an expectation value of $\hat{x}$ varies in time slowly
and put a certain special initial conditions.
Then, they show that the expectation value of $\hat{x}$ obeys the following equation:
\begin{equation}
\ddot{x}+\omega^{2}(1-\cos\Omega_{N}\tau)\sin x=0.
\label{equation-of-motion-x-recoil}
\end{equation}
Moreover, they obtain the effective Hamiltonian,
from which we can derive the above equation of motion, as follows:
\begin{equation}
H
=
\frac{1}{2}\dot{x}^{2}
-
\omega^{2}\cos x
+
\frac{\omega^{2}}{2}
[
\cos(x+\Omega_{N}\tau)
+
\cos(x-\Omega_{N}\tau)
].
\label{effective-Hamiltonian-x-recoil}
\end{equation}

In Eqs.~(\ref{equation-of-motion-x-recoil}) and (\ref{effective-Hamiltonian-x-recoil}),
we set $x=k_{f}\langle \hat{x} \rangle$ and $\tau=\Omega_{0}t$.
The new angular frequencies $\omega$ and $\Omega_{N}$ are quantities
constructed from $\omega_{a}$, $\omega_{f}$ and $\Omega_{0}$,
and they satisfy the relation
$\omega\propto\Omega_{N}^{-1}$.
The Hamiltonian given by Eq.~(\ref{effective-Hamiltonian-x-recoil}) represents
a particle moving in the field of three plane waves,
$\cos x$, $\cos (x+\Omega_{N}\tau)$ and $\cos (x-\Omega_{N}\tau)$.
It is widely known that this Hamiltonian induces chaotic dynamics.
We emphasize that Prants {\it et al.}'s results in Ref.~\cite{Prants2002}
are derived under the rotating-wave approximation.
In Ref.~\cite{Chotorlishvili2008},
Chotorlishvili and Toklikishvili generalize the Hamiltonian in Eq.~(\ref{extended-Hamiltonian-JCM-recoil})
for a three level optical atom and discuss its chaotic dynamics.

In Refs.~\cite{Milonni1983,Prants2002,Chotorlishvili2008},
the interactions between the atom and the cavity field are discussed under the semiclassical approximations.
In general,
it is very difficult to exactly solve the interaction between the two-level atom and the cavity field
as a fully quantum mechanical model without the rotating-wave approximation.
Hence,
we cannot predict whether or not actual experiments of the cavity QED
reproduce the stable results of quasiperiodicity and scale invariance.
This remains to be solved in future.

In this paper, we investigate some remarkable properties of trajectories
of the Bloch vector,
which is governed by the thermal JCM.
From careful observations of the quasiperiodic behaviour of the Bloch vector
under the thermal JCM,
we obtain novel and interesting facts,
such as
scale invariance and relation with the Diophantine approximation.
Throughout the latter half of this paper,
to examine quasiperiodic trajectories of the thermal Bloch vector precisely,
we borrow some useful concepts from the number theory,
which is a branch of the pure mathematics.
In Sec.~\ref{section-scale-invariance},
we introduce a knowledge about uniform distribution of real sequences.
In Sec.~\ref{section-perturbative-approach},
we utilize methods for obtaining rational approximation of irrational numbers
by means of their expressions of the continued fractions.

An appearance of Shor's algorithm lets the quantum information theory attract
many researchers' attention.
We remember that Shor's algorithm is a quantum algorithm for solving integer factorization and
discrete logarithm problems efficiently.
These problems are regarded as important topics in the field of the number theory and the cryptography.
Hence, progress of quantum information theory gives us new connections
between ideas in the number theory and the quantum mechanics.

Although the JCM was proposed and studied by researchers of quantum optics,
it has become familiar to those of quantum information theory.
The JCM has given various information and understandings on quantum mechanics to physicists.
The authors expect that we can bring out a great new variety of knowledge furthermore from the JCM,
especially, in connection with both discrete mathematics and quantum theory.

\appendix
\section{\label{section-appendix-A}
Physical transient spectra of the atom in the cavity}
In this section, we consider spectra of the atom,
which develops according to the Jaynes-Cummings interaction with the cavity field.
We discuss whether or not we can detect quasiperiodicity in the spectral analyses.
Moreover, to understand physical meanings of the discrete plots of the trajectories of the Bloch vector,
we investigate difference between continuous and discrete Fourier transforms for the atomic fluorescence.
At the end of this section,
to confirm the scale invariance for the discrete plots of the trajectories of the Bloch vector,
we examine histograms of the samples of the atomic fluorescence taken at a constant time interval.

First, we regard the time variable $t$ as continuous one,
and we define the physical transient spectrum of the radiation emitted by the atom as
\begin{equation}
S(\tilde{\omega})
=
2
\Gamma
\int_{-\infty}^{T}dt_{1}
\int_{-\infty}^{T}dt_{2}
e^{-(\Gamma-i\tilde{\omega})(T-t_{1})-(\Gamma+i\tilde{\omega})(T-t_{2})}
\langle\psi(t_{1})|
\sigma_{+}\sigma_{-}
|\psi(t_{2})\rangle,
\label{physical-transient-spectrum-0}
\end{equation}
where $\Gamma^{-1}$ represents the filter's response time
and $T$ represents the time at which the measurement takes place
\cite{Eberly1977,Sanchez-Mondragon1983,Gea-Banacloche1988}.

The reason why the physical transient spectrum
$S(\tilde{\omega})$ is given by Eq.~(\ref{physical-transient-spectrum-0}) is as follows.
First, because the output of the filter depends on past and current inputs
but not future ones,
the filter has to obtain causality.
Second, we require that $\Gamma^{-1}$ corresponds to the passband width.
Third, we let the filter extract an angular frequency component
from the input as the response.
Putting these requirements together,
we obtain a proper response of an actual filter as
\begin{equation}
h(t,\tilde{\omega},\Gamma)\propto\theta(t)e^{-(\Gamma+i\tilde{\omega})t}.
\end{equation}
Thus, we obtain Eq.~(\ref{physical-transient-spectrum-0}).

From now on, we calculate the physical transient spectrum of the atom evolving
under the thermal JCM according to Eq.~(\ref{physical-transient-spectrum-0}).
At first, we rewrite Eq.~(\ref{physical-transient-spectrum-0})
for cases where the state of the atom is described with a density operator,
\begin{eqnarray}
S(\tilde{\omega})
&=&
2\Gamma
\int_{-\infty}^{T}dt_{1}
\int_{-\infty}^{T}dt_{2}
e^{-(\Gamma-i\tilde{\omega})(T-t_{1})-(\Gamma+i\tilde{\omega})(T-t_{2})} \nonumber \\
&&
\quad
\times
\mbox{Tr}_{\mbox{\scriptsize P}}
[U(t_{2})\rho_{\mbox{\scriptsize AP}}(0)U^{\dagger}(t_{1})\sigma_{+}\sigma_{-}],
\end{eqnarray}
where $\rho_{\mbox{\scriptsize AP}}(0)=\rho_{\mbox{\scriptsize A}}(0)\otimes\rho_{\mbox{\scriptsize P}}$.

Here, as discussed in Sec.~\ref{section-trajectory-Bloch-vector},
we assume the initial state of the atom to be in the form
\begin{eqnarray}
\rho_{\mbox{\scriptsize A}}(0)
&=&
\frac{1}{2}
(|0\rangle_{\mbox{\scriptsize A}}+|1\rangle_{\mbox{\scriptsize A}})
({}_{\mbox{\scriptsize A}}\langle 0|+{}_{\mbox{\scriptsize A}}\langle 1|) \nonumber \\
&=&
\frac{1}{2}
\left(
\begin{array}{cc}
1 & 1 \\
1 & 1
\end{array}
\right).
\end{eqnarray}
Moreover, we assume that the initial state of the cavity field obeys the Bose-Einstein statistic
as Eq.~(\ref{field-initial-state}).
Then, we obtain
\begin{equation}
\sigma_{-}U(t_{2})\rho_{\mbox{\scriptsize A}}(0)\otimes\rho_{\mbox{\scriptsize P}}
U^{\dagger}(t_{1})\sigma_{+}
=
\frac{1}{2}
\left(
\begin{array}{cc}
0 & 0 \\
0 & \eta
\end{array}
\right),
\end{equation}
where
\begin{equation}
\eta
=
[
u_{00}(t_{2})
+
u_{01}(t_{2})
]
\rho_{\mbox{\scriptsize P}}
[
u^{\dagger}_{00}(t_{1})
+
u^{\dagger}_{01}(t_{1})
],
\end{equation}
and $u_{00}$ and $u_{01}$ are given by Eq.~(\ref{elements-time-evolution-op}).

Then calculating the partial traces of the operators, we obtain
\begin{eqnarray}
\lefteqn{\mbox{Tr}_{\mbox{\scriptsize P}}[u_{00}(t_{2})\rho_{\mbox{\scriptsize P}}u^{\dagger}_{00}(t_{1})]}
\nonumber \\
&=&
\sum_{n=0}^{\infty}
[
\cos(\sqrt{\tilde{D}(n)+g^{2}}t_{2})
+
\frac{i}{2}\Delta\omega
\frac{\sin(\sqrt{\tilde{D}(n)+g^{2}}t_{2})}{\sqrt{\tilde{D}(n)+g^{2}}}
] \nonumber \\
&&\quad
\times
(1-e^{-\beta\hbar\omega})e^{-\beta\hbar\omega n} \nonumber \\
&&\quad
\times
[
\cos(\sqrt{\tilde{D}(n)+g^{2}}t_{1})
-
\frac{i}{2}\Delta\omega
\frac{\sin(\sqrt{\tilde{D}(n)+g^{2}}t_{1})}{\sqrt{\tilde{D}(n)+g^{2}}}
], \nonumber \\
\lefteqn{\mbox{Tr}_{\mbox{\scriptsize P}}[u_{01}(t_{2})\rho_{\mbox{\scriptsize P}}u^{\dagger}_{01}(t_{1})]}
\nonumber \\
&=&
\sum_{n=1}^{\infty}
g^{2}n(1-e^{-\beta\hbar\omega})
\frac{\sin(\sqrt{\tilde{D}(n)}t_{2})}{\sqrt{\tilde{D}(n)}}
e^{-\beta\hbar\omega n}
\frac{\sin(\sqrt{\tilde{D}(n)}t_{1})}{\sqrt{\tilde{D}(n)}},
\end{eqnarray}
\begin{equation}
\mbox{Tr}_{\mbox{\scriptsize P}}[u_{00}(t_{2})\rho_{\mbox{\scriptsize P}}u^{\dagger}_{01}(t_{1})]
=
0,
\end{equation}
where $\tilde{D}(n)$ is given by Eq.~(\ref{definition-tilde-D}).
Here, as discussed in Sec.~\ref{section-trajectory-Bloch-vector},
we set $\Delta\omega=0$,
and we obtain
\begin{eqnarray}
\lefteqn{\mbox{Tr}_{\mbox{\scriptsize P}}[u_{00}(t_{2})\rho_{\mbox{\scriptsize P}}u^{\dagger}_{00}(t_{1})]}
\nonumber \\
&=&
(1-e^{-\beta\hbar\omega})
\sum_{n=1}^{\infty}
\cos(|g|\sqrt{n}t_{2})\cos(|g|\sqrt{n}t_{1})
e^{-\beta\hbar\omega (n-1)}, \nonumber \\
\lefteqn{\mbox{Tr}_{\mbox{\scriptsize P}}[u_{01}(t_{2})\rho_{\mbox{\scriptsize P}}u^{\dagger}_{01}(t_{1})]}
\nonumber \\
&=&
(1-e^{-\beta\hbar\omega})
\sum_{n=1}^{\infty}
\sin(|g|\sqrt{n}t_{2})\sin(|g|\sqrt{n}t_{1})
e^{-\beta\hbar\omega n}.
\end{eqnarray}
Thus, we arrive at
\begin{eqnarray}
&&
\mbox{Tr}
[U(t_{2})\rho_{\mbox{\scriptsize AP}}(0)U^{\dagger}(t_{1})\sigma_{+}\sigma_{-}] \nonumber \\
&=&
\frac{1}{2}
(1-e^{-\beta\hbar\omega})
\sum_{n=1}^{\infty}
[
\cos(|g|\sqrt{n}t_{2})\cos(|g|\sqrt{n}t_{1})
e^{-\beta\hbar\omega (n-1)} \nonumber \\
&&\quad
+
\sin(|g|\sqrt{n}t_{2})\sin(|g|\sqrt{n}t_{1})
e^{-\beta\hbar\omega n}].
\label{trace-atomic-signal-0}
\end{eqnarray}

Here, we set $\hbar=1$.
Then, replacing $\beta\omega$ and $|g|t$ with $\beta$ and $t$,
we substitute Eq.~(\ref{trace-atomic-signal-0}) into Eq.~(\ref{physical-transient-spectrum-0}).
With slightly tough calculations, we obtain
\begin{eqnarray}
S(\tilde{\omega})
&=&
\Gamma
\sum_{n=1}^{\infty}
\frac{e^{-(n+1)\beta}}{2[n+(\Gamma-i\tilde{\omega})^{2}][n+(\Gamma+i\tilde{\omega})^{2}]} \nonumber \\
&&\quad
\times
\Biggl[
(e^{2\beta}-1)(n+\Gamma^{2}+\tilde{\omega}^{2}) \nonumber \\
&&\quad
+
(e^{\beta}-1)^{2}
[(\Gamma^{2}+\tilde{\omega}^{2}-n)\cos(2\sqrt{n}T) \nonumber \\
&&\quad
+
2\Gamma\sqrt{n}\sin(2\sqrt{n}T)]
\Biggr].
\label{physical-transient-spectrum-1}
\end{eqnarray}
In the above equation, we assume that
$\Gamma$ and $\tilde{\omega}$ are in units of $\omega$
and $T$ is in units of $|g|^{-1}$.

\begin{figure}
\begin{center}
\includegraphics[scale=1.0]{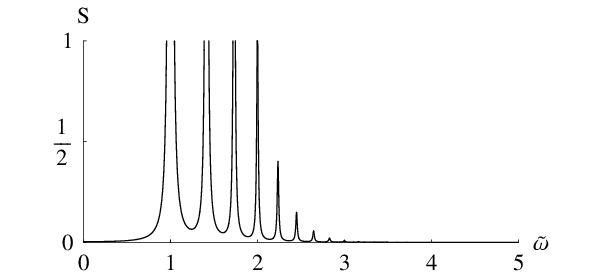}
\end{center}
\caption{The spectrum $S(\tilde{\omega})$ for $T=100.0$, $\Gamma=0.01$ and $\beta=1.0$.
We can observe peaks at $\tilde{\omega}=1,\sqrt{2},\sqrt{3},2,\sqrt{5},\sqrt{6},...$.}
\label{Figure12}
\end{figure}

Looking at Eq.~(\ref{physical-transient-spectrum-1}),
we notice
$S(-\tilde{\omega})=S(\tilde{\omega})$
and
$S^{\dagger}(\tilde{\omega})=S(\tilde{\omega})$.
We show the spectra $S(\tilde{\omega})$ of $\Gamma T=1$,
$T=100.0$ and $\beta=1.0$
in Fig.~\ref{Figure12}.
In Fig.~\ref{Figure12},
we can observe peaks of $\tilde{\omega}=\sqrt{n}$ for $n=1,2,3,...$.
These are manifestation of the quasiperiodicity of the system.
Moreover, from numerical calculations,
we can verify the following, at ease.
If $\beta$ becomes smaller,
that is,
if the temperature becomes higher,
the number of peaks of $S(\tilde{\omega})$ increases.

Here, let us think about the sampling theorem.
The sampling theorem tells us that if a signal has an upper angular frequency limit of $\tilde{\omega}$,
then we need sample points at time intervals being equal to or less than $\pi/\tilde{\omega}$
for enabling us to reconstruct the original signal.
Thus, to obtain information about the component of $\tilde{\omega}=\sqrt{n}$,
we have to take samples of the signal at time interval
being equal to or less than $\Delta t=\pi/\sqrt{n}$.
On the other hand, from the discussions about the scale invariance
and the pseudorandom sequence uniformly distributed for $[0,2\pi)$ in Sec.~\ref{section-scale-invariance},
we obtain $\pi<\Delta t<2\pi$ in Eq.~(\ref{condition- pseudorandom-Delta-t-0}).
Thus, for discretizing the dynamics of the Bloch vector,
we have to set $\pi<\Delta t<2\pi$.
In fact, in Figs.~\ref{Figure02}, \ref{Figure03}, \ref{Figure04}, \ref{Figure05}, \ref{Figure06}
and \ref{Figure07}
in Sec.~\ref{section-trajectory-Bloch-vector},
we take $\Delta t=3.5$ to convert the continuous trajectory of the Bloch vector
to a discrete-time sequence.

From these discussions,
we understand that discretization of the dynamics of the Bloch vector
with the time interval $\pi<\Delta t<2\pi$
destroys information of all components of $\tilde{\omega}=\sqrt{n}$
for $n=1, 2, 3, ...$,
so that we cannot recover the original trajectory.
This implies that the discretization with $\pi<\Delta t<2\pi$
erases the history of the trajectory of the Bloch vector completely.
Because of this effect,
figures consisting of the discrete-time sequence of the Bloch vector acquire the scale invariance.

In the following, we discuss the Fourier transform of the discrete-time sequence
obtained along the trajectory of the Bloch vector with the time interval $\pi<\Delta t<2\pi$.
Let us define the discrete Fourier transform of the fluorescence of the atom in the cavity as follows:
\begin{equation}
c_{k}
=
\frac{1}{N^{2}}
\sum_{l_{1}=0}^{N-1}
\sum_{l_{2}=0}^{N-1}
e^{-i2\pi(l_{2}-l_{1})k/N}
\mbox{Tr}
[
U(l_{2}\Delta t)\rho_{\mbox{\scriptsize AP}}(0)U^{\dagger}(l_{1}\Delta t)
\sigma_{+}\sigma_{-}].
\label{discrete-Fourier-transform-0}
\end{equation}
Looking at the above equation,
we notice $c_{k+N}=c_{k}$,
so that we concentrate only on $\{c_{k}:k=0,1,...,N-1\}$.
Substituting Eq.~(\ref{trace-atomic-signal-0}) into Eq.~(\ref{discrete-Fourier-transform-0}),
with slightly tough calculations, we obtain
\begin{eqnarray}
c_{k}
&=&
\sum_{n=1}^{\infty}
\frac{1}{N^{2}[\cos(2k\pi/N)-\cos(\sqrt{n}\Delta t)]^{2}}
e^{-n\beta}\sinh(\frac{\beta}{2}) \nonumber \\
&&
\times
\Biggl[
\cosh(\frac{\beta}{2})
[(1-\cos(\frac{2k\pi}{N})\cos(\sqrt{n}\Delta t))
(1-\cos(2k\pi)\cos(\sqrt{n}N\Delta t)) \nonumber \\
&&
-
\sin(2k\pi)\sin(\frac{2k\pi}{N})\sin(\sqrt{n}\Delta t)\sin(\sqrt{n}N\Delta t)] \nonumber \\
&&
+
\sinh(\frac{\beta}{2})
[\cos(\frac{2k\pi}{N})-\cos(\sqrt{n}\Delta t)]
\cos(\sqrt{n}(N-1)\Delta t) \nonumber \\
&&
\times
[\cos(2k\pi)-\cos(\sqrt{n}N\Delta t)]
\Biggr].
\label{discrete-Fourier-transform-1}
\end{eqnarray}

\begin{figure}
\begin{center}
\includegraphics[scale=1.0]{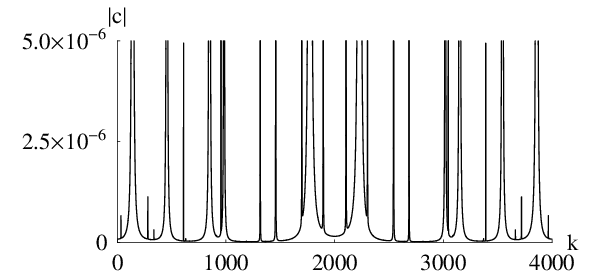}
\end{center}
\caption{The discrete Fourier transform of the Bloch vector
given by Eq.~(\ref{discrete-Fourier-transform-1})
for $\beta=1.0$, $\Delta t=3.5$ and $N=4000$.}
\label{Figure13}
\end{figure}

\begin{figure}
\begin{center}
\includegraphics[scale=1.0]{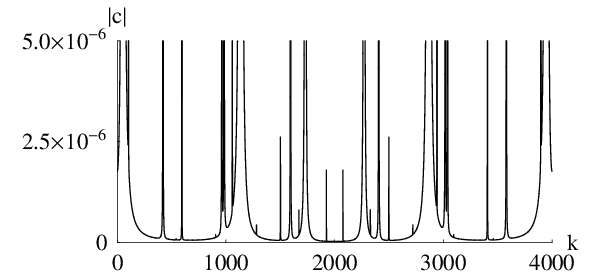}
\end{center}
\caption{The discrete Fourier transform of the Bloch vector
given by Eq.~(\ref{discrete-Fourier-transform-1})
for $\beta=1.0$, $\Delta t=4.5$ and $N=4000$.}
\label{Figure14}
\end{figure}

In Figs.~\ref{Figure13} and \ref{Figure14},
we plot $\{|c_{k}|:k=0,1,2,...,N-1\}$
for $\beta=1.0$, $\Delta t=3.5$, $N=4000$ and $\beta=1.0$, $\Delta t=4.5$, $N=4000$,
respectively.
Comparing Figs.~\ref{Figure13} and \ref{Figure14},
the discrete Fourier transforms of $\Delta t=3.5$ and $\Delta t=4.5$ are obviously different from each other.
This observation seems to insist that the scale invariance does not hold.
That is,
under a rescaling from $\Delta t=3.5$ to $\Delta t=4.5$,
the figure consisting of the discrete-time sequence of the Bloch vector changes.
However, this discussion is not true.

In fact, it is not significant to apply the Fourier analysis to behaviour of the discrete-time sequence
of the Bloch vector.
This is because we take a time interval $\pi<\Delta t<2\pi$ and we cannot recover entire information
about components of $\tilde{\omega}=\sqrt{n}$ for $n=1,2,...,N-1$.
That is to say,
a set of samples taken at time interval $\Delta t$ loses the past history of the trajectory.
When we apply the discrete Fourier transform to $N$ samples of a signal $f(t)$,
\begin{equation}
f(0), f(\Delta t), f(2\Delta t), ..., f((N-1)\Delta t),
\label{signal-discrete-sequence-0}
\end{equation}
we assume that they are arranged in chronological order.
However, the samples of the trajectory of the Bloch vector lose their past history completely,
so that $N$ samples cannot form a sequence of the chronological order shown
in Eq.~(\ref{signal-discrete-sequence-0}).

We can only regard the samples shown in Eq.~(\ref{signal-discrete-sequence-0})
as a set
\\
$\{f(n\Delta t):n=0,1,2,...,N-1\}$,
which loses the information about their order.
Here,
thinking about the scale invariance $\Delta t\to s\Delta t$ for $s>1$,
we want to show that we cannot distinguish between two sets of samples,
$\{f(n\Delta t):n=0,1,2,...,N-1\}$
and
$\{f(ns\Delta t):n=0,1,2,...,N-1\}$,
under the limit of $N\to\infty$.

Thus, we analyse the above two sets of the samples in the following way.
First, let us consider a histogram of the following set:
\begin{equation}
\{\mbox{Tr}
[U(l\Delta t)\rho_{\mbox{\scriptsize AP}}(0)U^{\dagger}(l\Delta t)\sigma_{+}\sigma_{-}]
:
l=0,1,2,...,N-1\}.
\label{signal-discrete-set-0}
\end{equation}
Second, we prepare two sets of Eq.~(\ref{signal-discrete-set-0})
for two different time intervals,
$\Delta t$ and $s\Delta t$.
If we cannot distinguish between their histograms under the limit of $N\to\infty$,
we can conclude that the scale invariance holds under $\Delta t\to s\Delta t$.

\begin{figure}
\begin{center}
\includegraphics[scale=1.0]{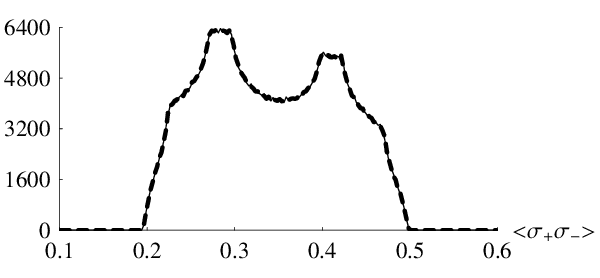}
\end{center}
\caption{The histograms of sets given by Eq.~(\ref{signal-discrete-set-0})
for $\Delta t=3.5$, $\beta=1.0$, $N=500\mbox{ }000$ and $\Delta t=4.5$, $\beta=1.0$, $N=500\mbox{ }000$
with a thin solid curve and a thick dashed curve, respectively.
In the graphs, we let the width of bins in the horizontal axis be equal to $0.0025$.}
\label{Figure15}
\end{figure}

In Fig.~\ref{Figure15},
we plot histograms of sets given by Eq.~(\ref{signal-discrete-set-0})
for
$\Delta t=3.5$, $\beta=1.0$, $N=500\mbox{ }000$
and
$\Delta t=4.5$, $\beta=1.0$, $N=500\mbox{ }000$
with a thin solid curve and a thick dashed curve,
respectively.
In Fig.~\ref{Figure15},
we let the width of bins in the horizontal axis be equal to $0.0025$.
Looking at Fig.~\ref{Figure15},
we can conclude that histograms of $\Delta t=3.5$ and $\Delta t=4.5$ are quite similar
and we can hardly distinguish between them.
Thus, we can confirm the scale invariance from Fig.~\ref{Figure15}.


\begin{thebibliography}{99}
%
\bibitem{Jaynes1963}
E.T.~Jaynes, F.W.~Cummings,
Comparison of quantum and semiclassical radiation theories with application to the beam maser,
Proc. IEEE 51(1)
(1963)
89--109.
%
\bibitem{Shore1993}
B.W.~Shore, P.L.~Knight,
The Jaynes-Cummings model,
J. Mod. Optics 40(7)
(1993)
1195--1238.
%
\bibitem{Louisell1973}
W.H.~Louisell,
Quantum Statistical Properties of Radiation,
John Wiley \& Sons, Inc., New York, U.S., 1973.
%
\bibitem{Barnett1997}
S.M.~Barnett, P.M.~Radmore,
Methods in Theoretical Quantum Optics,
Oxford University Press, Oxford, U.K., 1997.
%
\bibitem{Schleich2001}
W.P.~Schleich,
Quantum Optics in Phase Space,
Wiley-VCH, Berlin, 2001.
%
\bibitem{Cummings1965}
F.W.~Cummings,
Stimulated emission of radiation in a single mode,
Phys. Rev. 140(4A)
(1965)
A1051--A1056.
%
\bibitem{Eberly1980}
J.H.~Eberly, N.B.~Narozhny, J.J.~Sanchez-Mondragon,
Periodic spontaneous collapse and revival in a simple quantum model,
Phys. Rev. Lett. 44(20)
(1980)
1323--1326.
%
\bibitem{Rempe1987}
G.~Rempe, H.~Walther, N.~Klein,
Observation of quantum collapse and revival in a one-atom maser,
Phys. Rev. Lett. 58(4)
(1987)
353--356.
%
\bibitem{vonFoerster1975}
T.~von~Foerster,
A comparison of quantum and semi-classical theories of the interaction
between a two-level atom and the radiation field,
J. Phys. A: Math. Gen. 8(1)
(1975)
95--103.
%
\bibitem{Knight1982}
P.L.~Knight, P.M.~Radmore,
Quantum revivals of a two-level system driven by chaotic radiation,
Phys. Lett. A 90(7)
(1982)
342--346.
%
\bibitem{Knight1986}
P.L.~Knight,
Quantum fluctuations and squeezing in the interaction of an atom with a single field mode,
Phys. Scr. T12
(1986)
51--55.
%
\bibitem{Liu1992}
W.S.~Liu, P.~Tombesi,
Thermal photon distributions in the Jaynes-Cummings model,
Quantum Opt. 4(4)
(1992)
229--243.
%
\bibitem{Buck1981}
B.~Buck, C.V.~Sukumar,
Exactly soluble model of atom-phonon coupling showing periodic decay and revival,
Phys. Lett. A 81(2-3)
(1981)
132--135.
%
\bibitem{Arroyo-Correa1990}
G.~Arroyo-Correa, J.J.~Sanchez-Mondragon,
The Jaynes-Cummings model thermal revivals,
Quantum Opt. 2(6)
(1990)
409--421.
%
\bibitem{Chumakov1993}
S.M.~Chumakov, M.~Kozierowski, J.J.~Sanchez-Mondragon,
Analytical approach to the photon statistics
in the thermal Jaynes-Cummings model with an initially unexcited atom,
Phys. Rev. A 48(6)
(1993)
4594--4597.
%
\bibitem{Klimov1999}
A.B.~Klimov, S.M.~Chumakov,
Long-time behaviour of atomic inversion for the Jaynes-Cummings model in a strong thermal field,
Phys. Lett. A 264(2-3)
(1999)
100--102.
%
\bibitem{Fukuo1998}
T.~Fukuo, T.~Ogawa, K.~Nakamura,
Jaynes-Cummings model under continuous measurement:
weak chaos in a quantum system induced by unitarity collapse,
Phys. Rev. A 58(4)
(1998)
3293--3302.
%
\bibitem{Hioe1983}
F.T.~Hioe, H.-I.~Yoo, J.H.~Eberly,
Statistical analysis of long-term dynamic irregularity in an exactly soluble quantum mechanical model,
in:
J.~Chandra, A.C.~Scott (Eds.),
Coupled Nonlinear Oscillators,
North-Holland Publishing Co., Amsterdam, 1983, pp. 95--113.
%
\bibitem{Yoo1981}
H.-I.~Yoo, J.J.~Sanchez-Mondragon, J.H.~Eberly,
Non-linear dynamics of the fermion-boson model:
interference between revivals and the transition to irregularity,
J. Phys. A: Math. Gen. 14(6) (1981) 1383--1397.
%
\bibitem{Averbukh1989}
I.Sh.~Averbukh, N.F.~Perelman,
Fractional revivals:
universality in the long-term evolution of quantum wave packets
beyond the correspondence principle dynamics,
Phys. Lett. A 139(9) (1989) 449--453.
%
\bibitem{Parker1986}
J.~Parker, C.R.~Stroud,~Jr.,
Coherence and decay of Rydberg wave packets,
Phys. Rev. Lett. 56(7) (1986) 716--719.
%
\bibitem{Azuma2008}
H.~Azuma,
Dynamics of the Bloch vector in the thermal Jaynes-Cummings model,
Phys. Rev. A 77(6)
(2008)
063820.
%
\bibitem{Ott1993}
E.~Ott,
Chaos in Dynamical Systems,
Cambridge University Press, Cambridge, U.K., 1993.
%
\bibitem{Ruelle1984}
D.~Ruelle,
Strange attractors,
in:
P.~Cvitanovi{\'c} (Ed.),
Universality in Chaos,
2nd edn.,
Institute of Physics Publishing, Bristol, U.K., 1984,
pp.~37--48.
%
\bibitem{Goldstein2002}
H.~Goldstein, C.~Poole, J.~Safko,
Classical Mechanics,
3rd edn.,
Addison-Wesley, San Francisco, 2002.
%
\bibitem{Ostlund1983}
S.~Ostlund, D.~Rand, J.~Sethna, E.~Siggia,
Universal properties of the transition from quasi-periodicity to chaos in dissipative systems,
Physica D 8(3)
(1983)
303--342.
%
\bibitem{Glazier1988}
J.A.~Glazier, A.~Libchaber,
Quasi-periodicity and dynamical systems: an experimentalist's view,
IEEE Transactions on Circuits and Systems 35(7)
(July 1988)
790--809.
%
\bibitem{Arnold2006}
V.I.~Arnold, V.V.~Kozlov, A.I.~Neishtadt,
Mathematical Aspects of Classical and Celestial Mechanics,
3rd edn.,
Springer, Berlin, 2006.
%
\bibitem{Jordan2004}
T.F.~Jordan,
Steppingstones in Hamiltonian dynamics,
Am. J. Phys. 72(8)
(August 2004)
1095--1099.
%
\bibitem{Coppel2009}
W.A.~Coppel,
Number Theory: An Introduction to Mathematics,
2nd edn.,
Springer, Dordrecht, 2009.
%
\bibitem{Weyl1916}
H.~Weyl,
\"{U}ber die Gleichverteilung von Zahlen mod. Eins,
Mathematische Annalen 77(3)
(1916)
313--352.
%
\bibitem{Davenport1963}
H.~Davenport, P.~Erd\"{o}s, W.J.~LeVeque,
On Weyl's criterion for uniform distribution,
Michigan Math. J. 10(3)
(1963)
311--314.
%
\bibitem{Kuipers1974}
L.~Kuipers, H.~Niederreiter,
Uniform Distribution of Sequences,
John Wiley \& Sons, Inc., New York, U.S., 1974.
%
\bibitem{Walkden2011}
C.~Walkden,
Lecture notes on Ergodic Theory,
unpublished,
University of Manchester, 2011,
http://www.maths.manchester.ac.uk/{\textasciitilde}cwalkden/.
%
\bibitem{Niven1960}
I.~Niven, H.S.~Zuckerman,
An Introduction to the Theory of Numbers,
John Wiley \& Sons, Inc., New York, U.S., 1960.
%
\bibitem{Milonni1983}
P.W.~Milonni, J.R.~Ackerhalt, H.W.~Galbraith,
Chaos in the semiclassical $n$-atom Jaynes-Cummings model: failure of the rotating-wave approximation,
Phys. Rev. Lett. 50(13)
(1983)
966--969;
Phys. Rev. Lett. 51(12)
(1983)
1108.
%
\bibitem{Prants2002}
S.V.~Prants, M.~Edelman, G.M.~Zaslavsky,
Chaos and flights in the atom-photon interaction in cavity QED,
Phys. Rev. E 66(4)
(2002)
046222.
%
\bibitem{Chotorlishvili2008}
L.~Chotorlishvili, Z.~Toklikishvili,
Chaos in generalized
Jaynes-Cummings model,
Phys. Lett. A 372(16)
(2008)
2806--2815.
%
\bibitem{Eberly1977}
J.H.~Eberly, K.~W\'{o}dkiewicz,
The time-dependent physical spectrum of light,
J. Opt. Soc. Am. 67(9)
(1977)
1252--1261.
%
\bibitem{Sanchez-Mondragon1983}
J.J.~Sanchez-Mondragon, N.B.~Narozhny, J.H.~Eberly,
Theory
of spontaneous-emission line shape in an ideal cavity,
Phys. Rev. Lett. 51(7)
(1983)
550--553.
%
\bibitem{Gea-Banacloche1988}
J.~Gea-Banacloche, R.R.~Schlicher, M.S.~Zubairy,
Emission spectra of an atom in a cavity in the presence of a squeezed vacuum,
Phys. Rev. A 38(7)
(1988)
3514--3521.
\end{thebibliography}
\end{document}